\documentclass[prl,amsmath,amssymb,reprint,superscriptaddress]{revtex4-1}
\usepackage{hyperref}
\usepackage{graphicx}
\usepackage{braket}
\usepackage[subrefformat=parens,labelformat=parens,caption=false]{subfig}
\usepackage[inline]{enumitem}
\usepackage{xcolor}
\usepackage{soul}
\newcommand{\sites}{l}
\newcommand{\Sites}{L}

\begin{document}
\title{Many-Body Spin Echo}
\newcommand{\RegensburgUniversity}{Institut f\"ur Theoretische Physik, 
Universit\"at Regensburg, D-93040 Regensburg, Germany}
\newcommand{\LiegeUniversity}{Institute of Theoretical Physics, Liege University, Liege, Belgium}
\author{Thomas Engl}
\email{Thomas.Engl@physik.uni-regensburg.de}
\affiliation{\RegensburgUniversity}
\author{Juan Diego Urbina}
\affiliation{\RegensburgUniversity}
\author{Klaus Richter}
\affiliation{\RegensburgUniversity}
\author{Peter Schlagheck}
\affiliation{\LiegeUniversity}
\begin{abstract}
We predict a universal echo phenomenon present in the time evolution of many-body states of interacting quantum systems described by Fermi-Hubbard models. It consists of the coherent revival of transition probabilities echoing a sudden flip of the spins that, contrary to its single-particle (Hahn) version, is {\it not} dephased by interactions or spin-orbit coupling. The many-body spin echo signal has a universal shape independent of the interaction strength, and an amplitude and sign depending only on combinatorial relations between the number of particles and the number of applied spin flips. Our analytical predictions, based on semiclassical interfering amplitudes in Fock space associated with chaotic mean-field solutions, are tested against extensive numerical simulations confirming that the coherent origin of the echo lies in the existence of anti-unitary symmetries.   
\end{abstract}
%
\maketitle
%
%
Echoes such as the spin (or Hahn-) \cite{Hahn-echo}, mesoscopic \cite{mes-echo}, Loschmidt echo \cite{loschmidt_echo_introduction} and plasma wave echo \cite{plasmaecho} as well as time reversal focusing \cite{Fink} are among the fascinating quantum interference effects that have quickly become an important tool to characterize quantum coherence, stability of quantum dynamics, Anderson localization and the transition to classical behavior in quantum systems \cite{KlauderAnderson,loschmidt_rodolfo,hahn-decay,decoherence_hahn2,Cord,Cord_exp}. Echo phenomena also provide a way to gather information about many-body systems. Variations of the basic setup, the echo signal of a single degree of freedom coupled to a many-body system like a spin chain \cite{coupled_hahn_echo,spin-decoherence,spins_diamond} or of many atoms in an optical dipole trap \cite{hahn-echo_trapped_atoms}, are subject of present studies and can be used for measuring correlation functions and localization in many-body systems \cite{Knap_selfadvertisement1, Knap_selfadvertisement2}.

As most of the research on echo phenomena has focused on single-particle observables of many-body systems, interactions, acting as a coupling of a given particle with an external bath defined by the rest of the degrees of freedom, typically dephase the respective signals. Also, time-reversal symmetry breaking (crucial in the Hahn echo procedure) leads to dephasing of coherent effects like backscattering. It is therefore desirable to generate spin precession by an effective magnetic field which does {\it not} break time-reversal invariance as in systems with spin-orbit coupling (SOC) \cite{Rashba,Dresselhaus}. These ideas entered the cold-atoms community thanks to the realization of SOC in these systems, for which the spins can be macroscopically aligned \cite{macroscopic_singlet} and individually arranged on purpose \cite{soi_ultracold_atoms,soi_ultracold_atoms2,soi_ultracold_atoms3}. While these advances allow to model coherent effects in spintronic devices using optical lattices \cite{Lewenstein:2012,*ultracold-atoms_mimicking_condensed_matter-review}, one deals now with interacting systems where many-body effects beyond mean field approaches must be considered.
\begin{figure}[ttt]
\includegraphics[width=0.95\columnwidth]{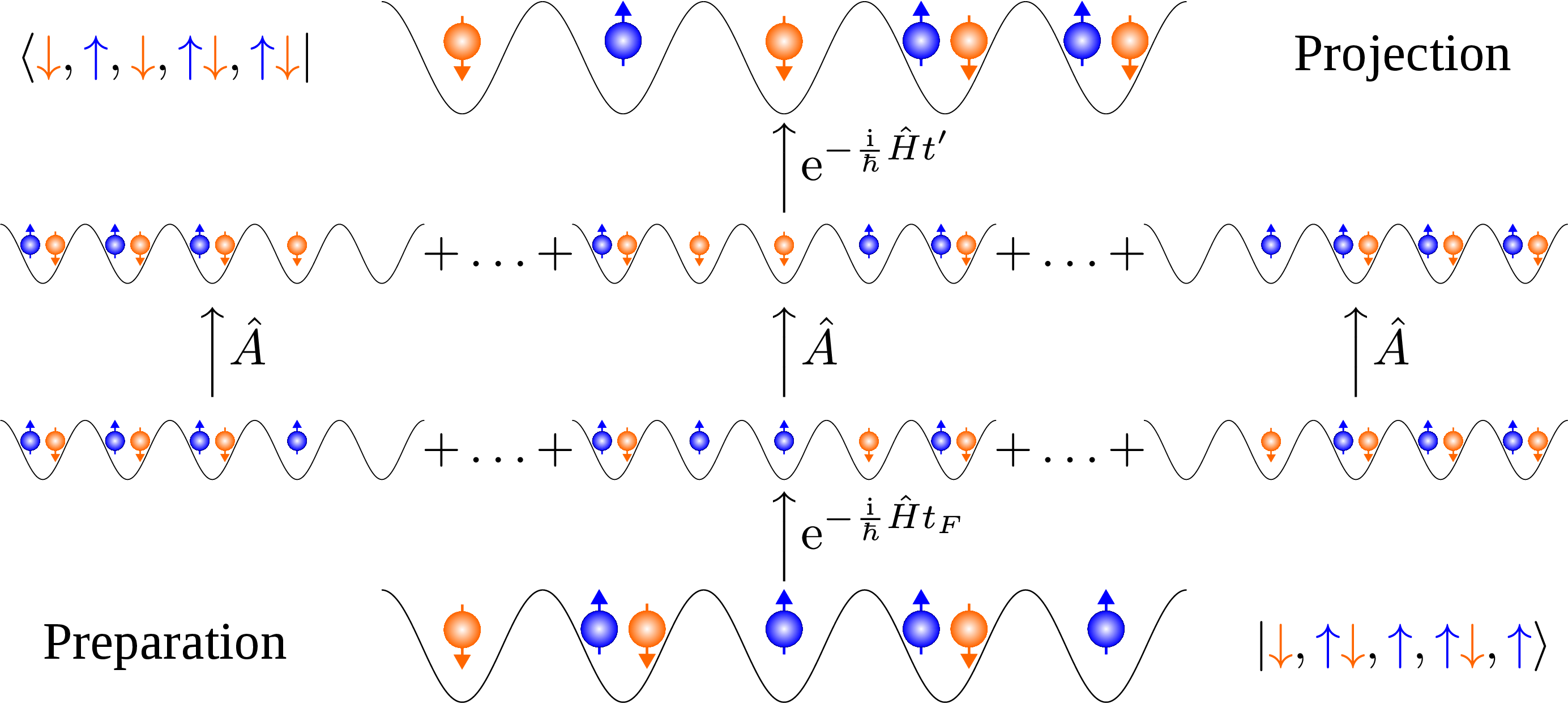}
\caption{\label{fig:fig1}Protocol of the many-body spin echo. A many-body initial Fock state (bottom line) evolves under the Hamiltonian $\hat{H}$, Eq.~(\ref{eq:quantum-hamiltonian}), describing both orbital and spin dynamics and the presence of interactions. At time $t_F$, when the system is generally in a superposition of Fock states, part of the spins are flipped through the spin flip operator $\hat{A}$ in Eq.~(\ref{eq:spin-flip_operator}). After further propagation for a time $t^{\prime}=t_F+\tau$ the occupations at each site are projectively measured (top line).}
\end{figure}

In this work we address the interplay between spin echo phenomena and many-body interference. Our main result is the prediction of a Many-Body Spin Echo (MBSE) effect akin to the Hahn echo but characteristic of interacting fermionic many-body systems with SOC and in the absence of magnetic fields. In contrast to single-particle echoes, inter-particle interactions -- as integral part of the many-body dynamics -- are {\it not} a source of decoherence of the MBSE signal. In the non-perturbative regime where SOC and hopping compete with the interaction terms, the MBSE signal  takes a universal form. Our semiclassical approach in terms of interfering paths in Fock space enables us to identify the coherent mechanism responsible for the MBSE as the constructive interference between amplitudes associated with classical mean-field solutions related by anti-unitary symmetries, and to provide analytical results for the amplitude and width of the echo signal. Its observation is in reach of state of the art experiments with fermionic cold atoms \cite{soi_ultracold_atoms,soi_ultracold_atoms2,soi_ultracold_atoms3}.

The setup of the MBSE is sketched in Fig.~\ref{fig:fig1}. We consider a system of $N$ itinerant interacting fermions on a lattice of $L$ sites in the presence of SOC. Initially, the system is prepared in a Fock state $\ket{\bf n}=\ket{n_{1\uparrow},n_{1\downarrow},\ldots,n_{L\uparrow},n_{L\downarrow}}$, where $n_{i \uparrow (\downarrow)} \in \{0,1\}$ denotes the occupation number with spin up (down) at the $i$th site. After the system has evolved for some time $t_F$, the spins of the particles at $M \leq L$ sites \footnote{The result will not depend on the choice of the sites. To simplify the notation we take the first $M$.} are suddenly reversed through the spin flip operator $\hat{A}=\prod_{\sites=1}^{M}\hat{A}_{\sites}$ with \footnote{Adding an additional phase factor ${\rm e}^{{\rm i}\chi}$ to the spin flip operator, $\hat{A}_{\sites}=\left(1-\hat{n}_{\sites,\uparrow}^{}\right)\left(1-\hat{n}_{\sites,\downarrow}^{}\right)+\hat{n}_{\sites,\uparrow}^{}\hat{n}_{\sites,\downarrow}^{}+{\rm e}^{{\rm i}\chi}\hat{c}_{\sites,\uparrow}^{\dagger}\hat{c}_{\sites,\downarrow}^{}+{\rm e}^{-{\rm i}\chi}\hat{c}_{\sites,\downarrow}^{\dagger}\hat{c}_{\sites,\uparrow}^{}$ corresponds to changing the Rashba phase $\varphi=\arg(\kappa_{\uparrow\downarrow})$ to $\varphi+\chi$.} 
\begin{equation}
\hat{A}_{\sites}=\left(1-\hat{n}_{\sites,\uparrow}^{}\right)\left(1-\hat{n}_{\sites,\downarrow}^{}\right)+\hat{n}_{\sites,\uparrow}^{}\hat{n}_{\sites,\downarrow}^{}
+\hat{c}_{\sites,\uparrow}^{\dagger}\hat{c}_{\sites,\downarrow}^{}+\hat{c}_{\sites,\downarrow}^{\dagger}\hat{c}_{\sites,\uparrow}^{}.
\label{eq:spin-flip_operator}
\end{equation}
The system is then further propagated for a time $t'=t_F+\tau$ and the probability to obtain the occupations ${\bf n}^\prime$,
\begin{equation}
P\left({\bf n}^\prime,{\bf n};\tau\right)=\left|\Braket{{\bf n}^\prime|{\rm e}^{-\frac{i}{\hbar}\hat{H}(t_F+\tau)}\hat{A}{\rm e}^{-\frac{i}{\hbar}\hat{H}t_F}|\bf n}\right|^2,
\label{eq:timeev}
\end{equation}
is measured as a function of the time mismatch $\tau$. Here, the many-body Hamiltonian for a Fermi-Hubbard ring with on-site energies $\epsilon_{l}$, interactions $U$ and nearest-neighbor hopping $\kappa_{\sigma\sigma^\prime}$ is 
\begin{equation}
\begin{split}
\hat{H}=\sum\limits_{\sites=1,L}\left[\epsilon_{\sites}\left(\hat{c}_{\sites\uparrow}^\dagger\hat{c}_{\sites\uparrow}^{}+\hat{c}_{\sites\downarrow}^\dagger\hat{c}_{\sites\downarrow}^{}\right)+U\hat{c}_{\sites\uparrow}^\dagger\hat{c}_{\sites\downarrow}^\dagger\hat{c}_{\sites\downarrow}^{}\hat{c}_{\sites\uparrow}^{}\right]& \\
+\sum\limits_{\sites=1}^{L-1}\sum\limits_{\sigma,\sigma^\prime}\kappa_{\sigma\sigma^\prime}\left(\hat{c}_{\sites\sigma}^\dagger\hat{c}_{\sites+1\sigma^\prime}+\hat{c}_{\sites+1\sigma}^\dagger\hat{c}_{\sites\sigma^\prime}\right)&.
\end{split}
\label{eq:quantum-hamiltonian}
\end{equation}
The term $\kappa_{\sigma\sigma^\prime}$ represents spatially homogeneous SOC, $\kappa_{\uparrow,\downarrow}=\kappa_{\downarrow,\uparrow}^{\ast}=\alpha{\rm e}^{{\rm i}\varphi}$, and hopping $\kappa_{\sigma,\sigma}=J$.

\begin{figure}[ttt]
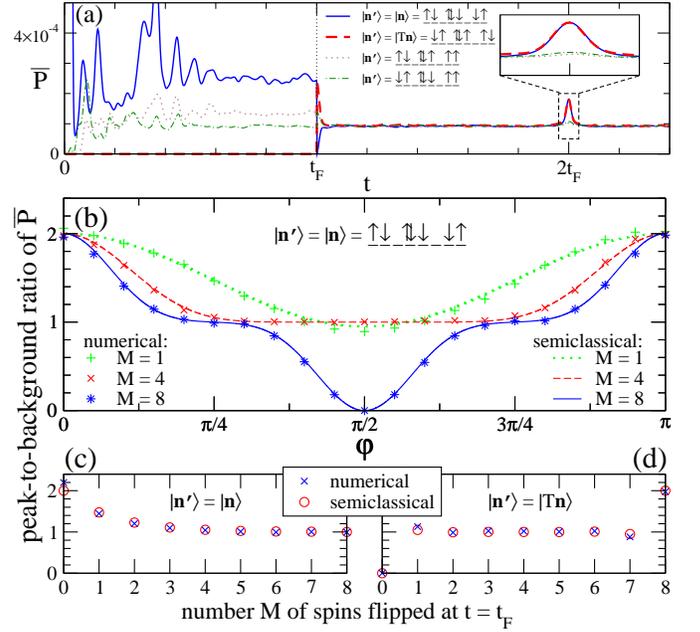

\includegraphics[width=0.49\textwidth]{figs/Spinecho.eps}
\includegraphics[width=0.49\textwidth]{figs/Peakheights.eps}
\caption{Many-body spin echo in a Fermi-Hubbard system. a) Average $\overline{P}$ of the probability $P$ in Eq.~(\ref{eq:timeev}) to obtain the final Fock state $|{\bf n}^{\prime}\rangle=|{\bf X}{\bf n}\rangle$ following the protocol in Fig.~\ref{fig:fig1}. We show $\overline{P}$ for ${\bf X}={\rm id},{\bf T}$ corresponding to the final measurement of the initial occupations $|{\bf n}\rangle$ (id), of their image under spin flip (${\bf T}$) as well as for two other final states that are uncorrelated with $|{\bf n}\rangle$. Coherent effects for ${\bf X}={\rm id},{\bf T}$ before spin flip ($t \le t_{\rm F}$) are lost after application of the spin-flip operator $\hat{A}$ at time $t_{\rm F}$ and further propagation ${\rm exp}[-i\hat{H}(t_{\rm F}+\tau)/\hbar]$. At $\tau=0$ and for ${\bf X}={\rm id},{\bf T}$, however, the probability echoes the initial state with an enhancement not present for other final states, a consequence of Eq.~(\ref{eq:peak}) here numerically confirmed for $N=7$ particles in $L=8$ sites and real SOC ($\kappa_{\uparrow,\downarrow}=\kappa_{\downarrow,\uparrow}^{\ast}=\alpha{\rm e}^{{\rm i}\varphi}$ with $\varphi=0$ in Eq.~(\ref{eq:quantum-hamiltonian})). b) According to Eqs.~(\ref{eq:echo-signal_si},\ref{eq:echo-signal_si2}) the contrast (peak/background ratio of $\overline{P}$) has a universal dependence on $N$, $L$, number of sites $M$ where the spins are flipped at $t=t_{\rm F}$, and $\varphi$ as symmetry parameter in Eq.~(\ref{eq:anti_unit}). Panels c) and d) show the dependence of the contrast on $M$ for ${\bf X}={\rm id}$ and ${\bf X}={\bf T}$ when $\varphi=\pi/4$. For the simulations we consider a Fermi-Hubbard ring with $N=7,L=8,\alpha=J/5,U=J$ and $\epsilon_{l} \in [0, 2J]$ in Eq.~(\ref{eq:quantum-hamiltonian}). We use $|{\bf n}\rangle=|\uparrow,\downarrow,0,\uparrow\downarrow,\downarrow,0,\downarrow,\uparrow\rangle$ and $t_{\rm F}=25\hbar/J$  but the results do not depend on these choices.     
}  
\label{fig:fig2}
\end{figure}

We denote by $\overline{P}\left(X{\bf n},{\bf n};\tau\right)$ the average (over random energies $\epsilon_{l}$) of the probability $P$ in Eq.~(\ref{eq:timeev}) to measure ${\bf n}^{\prime}={\bf X}{\bf n}$ after the total propagation time $t_{{\rm F}}+t'=2t_{{\rm F}}+\tau$. For $\ket{{\bf n}^\prime}=\ket{{\bf n}}$, one has ${\bf X}={\rm id}={\rm diag}({\bf 1}_{2},\ldots,{\bf 1}_{2})$ with ${\bf 1}_{2}$ the $2\times 2$ unit matrix, while for the final state being the spin inverse of the initial state, $\ket{{\bf n}^\prime}=\ket{{\bf T}{\bf n}}$, one has ${\bf X}={\bf T}={\rm diag}({\boldsymbol\sigma}_x,\ldots,{\boldsymbol\sigma}_x)$ with ${\boldsymbol\sigma}_x$ the $x$-Pauli matrix. Our analytical predictions of the key features of $\overline{P}$ are shown in Fig.~\ref{fig:fig2} together with their numerical confirmation. They support our claim about the existence of a (robust, non-perturbative and universal) coherent echo mechanism due to many-body quantum interference.   

In the following we explain how the analytical results for $\overline{P}$ that follow from Eqs.~(\ref{eq:echo-signal_tr},\ref{eq:echo-signal_si}, and \ref{eq:peak}), and are depicted in Fig.~\ref{fig:fig2}, are obtained within the semiclassical approximation for the microscopic path integral propagator of discrete fermionic quantum fields derived in \cite{fermionic_propagator}. There, the classical limit (in the sense of $N \gg 1$) was shown to be a Hamiltonian theory for a classical, complex multicomponent field ${\boldsymbol\psi}=(\psi_{1 \uparrow},\psi_{1 \downarrow},\ldots,\psi_{L \uparrow},\psi_{L \downarrow})$ with dynamics generated by Hamilton's equations
\begin{equation}
{\rm i}\hbar\frac{d}{ds}{\boldsymbol\psi}(s)=\frac{\partial H_{MF}\left({\boldsymbol\psi}^\ast,{\boldsymbol\psi}\right)}{\partial{\boldsymbol\psi}^\ast},
\label{eq:eom}
\end{equation}
in terms of the classical (mean-field like) Hamiltonian $H_{MF}$ \cite{fermionic_propagator} (whose explicit form will not be relevant in the following), and supplemented with boundary conditions at initial ($s=0$) and final time ($s=t$),
\begin{equation}
|\psi_{i \uparrow (\downarrow)}(s=0)|^{2}=n_{i \uparrow (\downarrow)}, {\rm \ \ }|\psi_{i \uparrow (\downarrow)}(s=t)|^{2}=n'_{i \uparrow (\downarrow)}.
\label{eq:bouncond}
\end{equation}
Equations~(\ref{eq:eom},\ref{eq:bouncond}) admit a discrete set of solutions ${\boldsymbol\psi}_{\gamma}(s)$ indexed by $\gamma$, and the field propagator is approximated by a coherent sum over interfering amplitudes \cite{fermionic_propagator}
\begin{equation}
\Braket{{\bf n}^\prime|{\rm e}^{-\frac{i}{\hbar}\hat{H}t}|\bf n}\simeq\sum_{\gamma:{\bf n}\to{\bf n}^{\prime}}\mathcal{A}_\gamma{\rm e}^{\frac{\rm i}{\hbar}R_{\gamma}}.
\label{eq:prop}
\end{equation}
In Eq.~(\ref{eq:prop}) the classical actions $R_{\gamma}=\int_{0}^{t}{\rm d}s\left[\hbar{\boldsymbol\theta}\cdot\dot{\bf I}-H_{MF}\left({\boldsymbol\psi}^\ast,{\boldsymbol\psi}\right)\right]$ and the semiclassical amplitudes $\mathcal{A}_\gamma$ depend both on the mean-field solutions $\psi_{j\sigma}(s)=\sqrt{I_{j\sigma}(s)}{\rm e}^{{\rm i}\theta_{j\sigma}(s)}$ (see Ref.~\cite{fermionic_propagator} for details). Since Eq.~(\ref{eq:eom}) generically displays chaotic behavior \cite{simon}, the BGS conjecture of quantum chaos predicts the emergence of universal signatures of quantum interference \cite{BGS} strengthened by averages respecting the symmetries of $\hat{H}$ in Eq.~(\ref{eq:quantum-hamiltonian}). Equations~(\ref{eq:timeev},\ref{eq:prop}) make such interferences (between many-body amplitudes) explicit through a coherent four-fold sum over mean-field solutions (paths)
\begin{eqnarray}
\label{eq:mbse_probability_sc_full}
&&P\left({\bf n}^\prime,{\bf n};\tau\right)\approx \\ && \sum\limits_{{\bf m}^{},{\bf m}^{\prime}}\sum\limits_{\substack{\gamma_{1}:{\bf n}\to{\bf m} \\ \gamma_3:{\bf n}\to{\bf m}^{\prime}}}^{}\sum\limits_{\substack{\gamma_{2}:{\bf F}^{(M)}{\bf m}\to{\bf n}^{\prime} \\ \gamma_{4}:{\bf F}^{(M)}{\bf m}^{\prime}\to{\bf n}^{\prime}}}^{}\mathcal{A}_{\gamma_1}\mathcal{A}_{\gamma_2}\mathcal{A}_{\gamma_3}^\ast\mathcal{A}_{\gamma_4}^\ast \exp\left[{\rm i} \Delta_{\gamma_{1}\gamma_{2}}^{\gamma_{3}\gamma_{4}}\right], \nonumber
\end{eqnarray}
with combined action differences 
\begin{equation}
\Delta_{\gamma_{1}\gamma_{2}}^{\gamma_{3}\gamma_{4}}=(R_{\gamma_1}+R_{\gamma_2}-R_{\gamma_3}-R_{\gamma_4})/\hbar.
\label{eq:Delta}
\end{equation} 
The action of the spin inversion operator $\hat{A}$ is incorporated in the boundary conditions of the mean-field solutions by multiplying the intermediate occupation vectors ${\bf m}$ and ${\bf m}^\prime$ by the $2L\times2L$ block-diagonal spin-flip matrix ${\bf F}^{(M)}={\rm diag}({\bf f}_{1},\ldots,{\bf f}_{M},{\bf f}_{M+1},\ldots,{\bf f}_{L})$ with ${\bf f}_{i=1,\ldots,M}={\boldsymbol \sigma}_{x}$ and ${\bf f}_{i=M+1,\ldots,L}={\bf 1}_2$.
\begin{figure}
\centering
\subfloat[\label{fig:diagonal}]{
\includegraphics[width=0.2\textwidth]{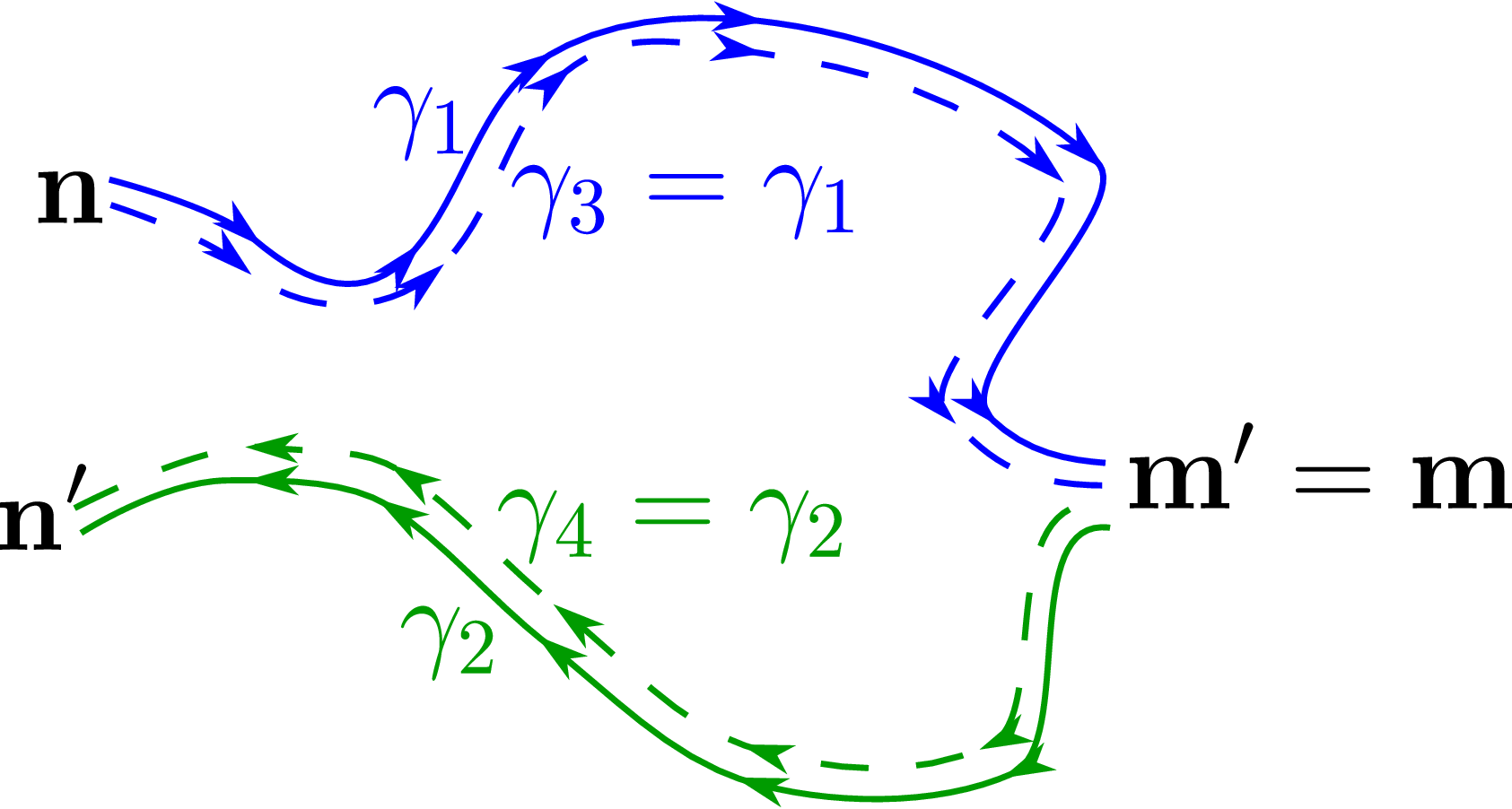}
}
\hspace{2em}
\subfloat[\label{fig:cbs_crossed_all}]{
\includegraphics[width=0.2\textwidth]{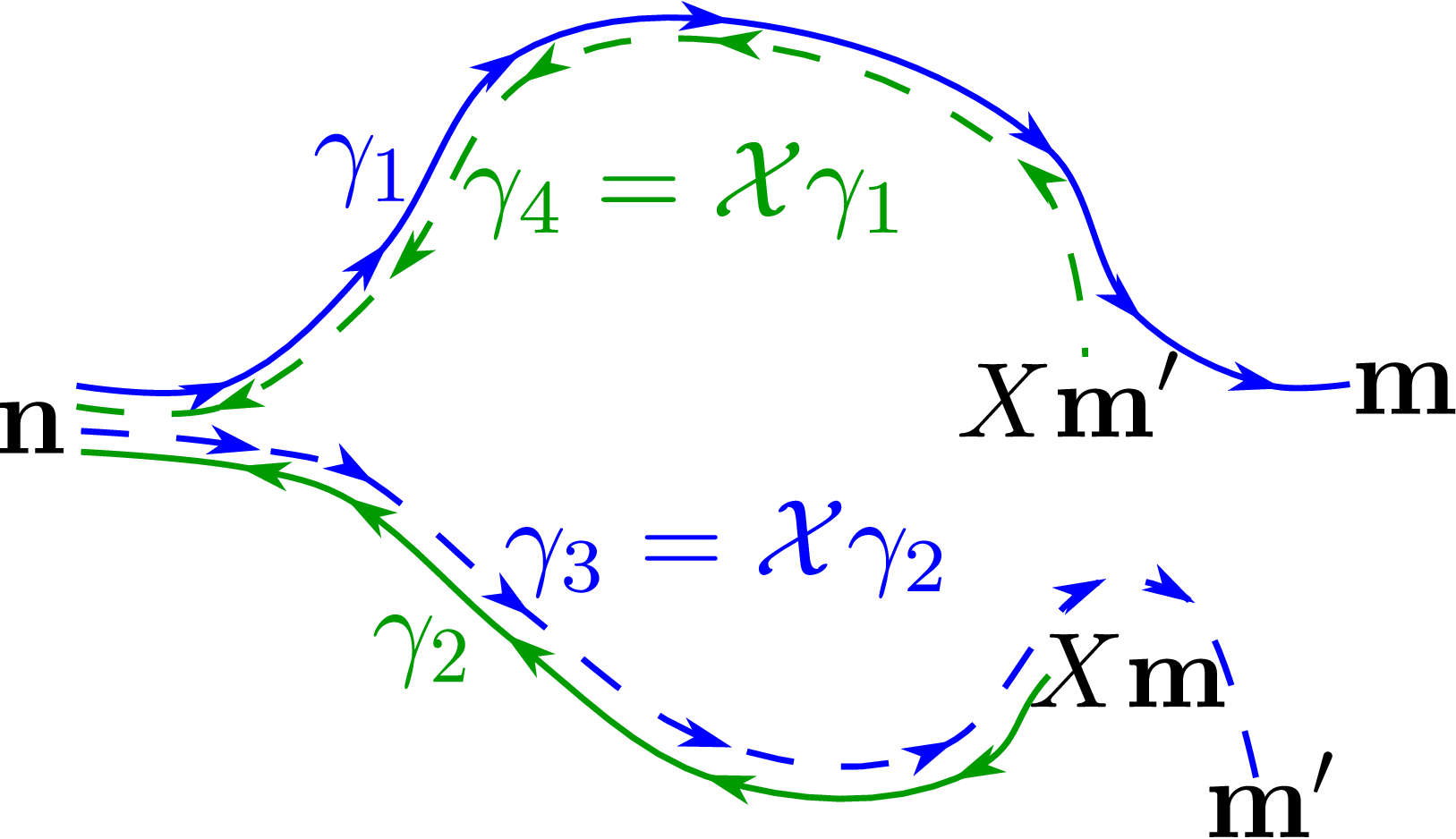}
} \\
\subfloat[\label{fig:qm_background}]{
\includegraphics[width=\columnwidth]{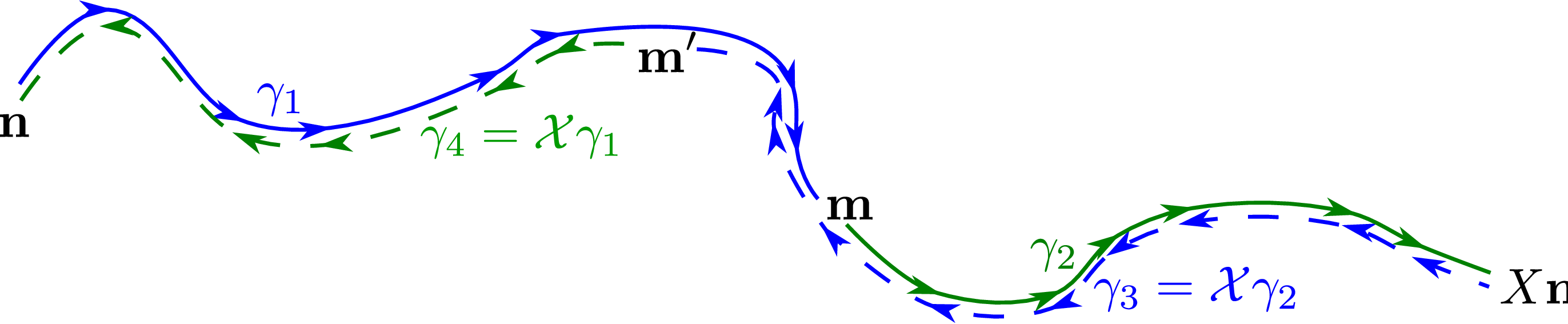}
}
\caption{\label{fig:leading_order_spin_echo}Leading order contributions $(\gamma,\mathcal{X}\gamma)$ to the averaged many-body probability, Eq.~(\ref{eq:mbse_probability_sc_full}), where either $\mathcal{X}=\mathcal{T}\mathcal{S}_{\varphi}$ and $X={\rm id}$ or $\mathcal{X}=\mathcal{T}$ and ${\bf X}={\bf T}$. \protect\subref{fig:diagonal} Diagonal approximation, \protect\subref{fig:cbs_crossed_all} echo contribution, \protect\subref{fig:qm_background} quantum correction to background.}
\end{figure}

Large action differences $\Delta_{\gamma_{1}\gamma_{2}}^{\gamma_{3}\gamma_{4}} \propto N \gg 1$ in the exponentials of Eq.~(\ref{eq:mbse_probability_sc_full}) give rise to fast oscillations, which generally cancel out upon average unless there are classical correlations between pairs of actions, determined solely by the anti-unitary symmetries of the Hamiltonian \eqref{eq:quantum-hamiltonian}. From all the possibilities $(\gamma,\gamma')$ of pairing, however, some require conditions on both intermediate occupations ${\bf m}$ and ${\bf m}^\prime$, and thus yield negligible contributions. Neglecting furthermore so-called loop contributions 
\cite{SR-pairs,quantumchaostransportKlaus,SR-pairs_dg3,3-encounter,loops_transport},
to get a correlated pair $(\gamma,\mathcal{X}\gamma)$ one can pair each trajectory $\gamma$ with itself, with its time reverse $\mathcal{T}\gamma$, 
\begin{equation}
\left(\begin{array}{c} {\boldsymbol\psi}_{\uparrow}^{(\mathcal{T}\gamma)}(t) \\ {\boldsymbol\psi}_{\downarrow}^{(\mathcal{T}\gamma)}(t) \end{array}\right)=\left(\begin{array}{c} \left({\boldsymbol\psi}_{\downarrow}^{(\gamma)}\right)^\ast(t_\gamma-t) \\ -\left({\boldsymbol\psi}_{\uparrow}^{(\gamma)}\right)^\ast(t_\gamma-t) \end{array}\right),
\label{eq:time_reverse_trajectory}
\end{equation}
or with the time reverse $\mathcal{T}\mathcal{S}_{\varphi}\gamma$ of its spin-reversed $\mathcal{S}_{\varphi}\gamma$, 
\begin{equation}
\left(\begin{array}{c} {\boldsymbol\psi}_{\uparrow}^{(\mathcal{S}_{\varphi}\gamma)}(t) \\ {\boldsymbol\psi}_{\downarrow}^{(\mathcal{S}_{\varphi}\gamma)}(t) \end{array}\right)=\left(\begin{array}{c} {\boldsymbol\psi}_{\downarrow}^{(\gamma)}(t){\rm e}^{-{\rm i}\varphi} \\ -{\boldsymbol\psi}_{\uparrow}^{(\gamma)}(t){\rm e}^{{\rm i}\varphi} \end{array}\right),
\label{eq:anti_unit}
\end{equation}
where ${\boldsymbol\psi}_{\uparrow/\downarrow}^{(\gamma)}$ is the vector containing all spin-up/spin-down components of the mean-field solution $\gamma$. As shown in Fig.~\ref{fig:leading_order_spin_echo} the resulting sets of correlated actions giving the leading order to the average probability $\overline{P}$ are 
\begin{enumerate*}[label=(\roman*)]
\item $\gamma_3=\gamma_1$ and $\gamma_4=\gamma_2$ (panel~\subref{fig:diagonal}),
\label{enum:diagonal}
\item $\gamma_4=\mathcal{T}\gamma_1$ and $\gamma_3=\mathcal{T}\gamma_2$ (panel~\subref{fig:cbs_crossed_all} for intermediate occupations with ${\bf m}\neq{\bf F}^{(M)}{\bf m}$ and panel~\subref{fig:qm_background} else), and
\label{enum:tr-pairing}
\item $\gamma_4=\mathcal{T}\mathcal{S}_{\varphi}\gamma_1$ and $\gamma_3=\mathcal{T}\mathcal{S}_{\varphi}\gamma_2$ (again panel~\subref{fig:cbs_crossed_all} for intermediate occupations with ${\bf m}\neq{\bf F}^{(M)}{\bf m}$ and panel~\subref{fig:qm_background} else).
\label{enum:si-pairing}
\end{enumerate*}
The pairing \ref{enum:diagonal}, also known as diagonal approximation \cite{da}, obviously requires ${\bf m}^\prime={\bf m}$ and yields the classical, incoherent ($\Delta_{\gamma_{1}\gamma_{2}}^{\gamma_{1}\gamma_{2}}=0$) contribution 
\begin{equation}
\overline{P}_{\rm cl}\left({\bf n}^{\prime},{\bf n};\tau\right)=\sum\limits_{\bf m}p_{{\rm cl}}({\bf n}^\prime,{\bf F}^{(M)}{\bf m};t_F+\tau)p_{{\rm cl}}({\bf m},{\bf n};t_F)
\label{eq:incoherent_contribution}
\end{equation}
meaning that the two propagation steps before and after the spin-flip are independent. In Eq.~(\ref{eq:incoherent_contribution}), $p_{{\rm cl}}({\bf m},{\bf n};t)$ is the averaged classical probability to obtain a set of occupations ${\bf m}$ after evolving the classical phase space distribution representing ${\bf n}$ for a time $t$ under mean-field dynamics \cite{fermionic_propagator}, as explicitely done for bosons in \cite{cbs_fock}. 

Quantum interference effects manifest themselves in deviations from this classical background. They are encoded in the echo signal ${\cal P}^{\rm echo}_{M,X}$, conveniently defined as
\begin{equation}
{\cal P}^{\rm echo}_{M,X}\left({\bf n};\tau\right):=\frac{\overline{P}\left(X{\bf n},{\bf n};\tau\right)}{\overline{P}_{\rm cl}\left(X{\bf n},{\bf n};\tau\right)}.
\label{eq:def_mbse-probability}
\end{equation}

The crosswise pairings \ref{enum:tr-pairing} and \ref{enum:si-pairing} giving the \emph{coherent} contributions, require that either the difference between the two propagation times is very small, $\tau\sim0$, (Fig.~\subref*{fig:cbs_crossed_all}) giving rise to the echo peak, or that the spin flips do not change the intermediate occupations, {\it i.e.}~${\bf m}={\bf F}^{(M)}{\bf m}$ (Fig.~\subref*{fig:qm_background}), which will describe the remnants of the usual transition probabilities without the spin-flips \footnote{For $M=0$ this yields the transition probability $P=p_{\rm cl}(1+\delta_{{\bf n}^{\prime},{\bf n}}+\left(-1\right)^{N}\delta_{{\bf n}^{\prime},{\bf F}^{(L)}{\bf n}})$ along the lines of the studies of coherent forward- and backscattering.~\cite{cbs_fock,fermionic_propagator}}. In this latter case, the sum is dominated by pairs of trajectories $\gamma_1,\gamma_2$ joining smoothly, such that when pairing \ref{enum:tr-pairing} $\gamma_4=\mathcal{T}\gamma_1$ and $\gamma_3=\mathcal{T}\gamma_2$ results in $\Delta_{\gamma_{1}\gamma_{2}}^{\gamma_{3}\gamma_{4}}\simeq \pi\hbar\sum_{l=1}^{L}(n_{l\uparrow}-n_{l\downarrow})$ and \ref{enum:si-pairing} $\gamma_4=\mathcal{T}\mathcal{S}_{\varphi}\gamma_1$ while $\gamma_3=\mathcal{T}\mathcal{S}_{\varphi}\gamma_2$ results in $\Delta_{\gamma_{1}\gamma_{2}}^{\gamma_{3}\gamma_{4}}\simeq 0$. For ergodic classical transition probabilities $p_{\rm cl}$, the echo probability for $\tau\nsim0$ is then given by the ratio
\begin{equation}
\frac{\mathcal{N}_{\rm e}^{(0)}}{\mathcal{N}}=\sum\limits_{k=0}^{\min\left(M,\left\lfloor \frac{N}{2} \right\rfloor\right)}{M \choose k}{2\left(\Sites-M\right) \choose N-2k}\bigg/{2\Sites \choose N}
\label{eq:number_double-occupancies}
\end{equation}
between the number $\mathcal{N}_{\rm e}^{(0)}$ of states ${\bf m}$ with the sites $1,\ldots,M$ either empty or doubly occupied, and the total number $\mathcal{N}$ of states ($\lfloor x \rfloor$ denotes integer part of $x$).

For $\tau\sim0$, on the other hand, the contribution \ref{enum:tr-pairing} requires that ${\bf m}^\prime={\bf T}{\bf F}^{(M)}{\bf m}$ and, more importantly, ${\bf n}^\prime={\bf T}{\bf n}$, where ${\bf T}={\bf F}^{(L)}$ is the matrix corresponding to full spin flip $M=L$. Since the sum~(\ref{eq:Delta}) of the action differences, obtained by noticing that each pair of trajectory has the same energy and plugging in the relations between them into the kinetic part of the action, 
\begin{equation}
\Delta_{\gamma_{1}\gamma_{2}}^{\mathcal{T}\gamma_{1}\mathcal{T}\gamma_{2}}=\pi\sum\limits_{\sites=1}^{\Sites}\left[n_{\sites\downarrow}-n_{\sites\uparrow}+m_{\sites\uparrow}-\left({\bf F}^{(M)}{\bf m}\right)_{\sites\uparrow}\right]
\label{eq:action_difference_echo-ti}
\end{equation}
is an integer multiple of $\pi$, each term in the sum over intermediate occupations contributes with a negative sign if the number of particles in the last $\Sites-M$ states is odd, and with a positive sign otherwise. Defining $\eta_{{\rm e(o)}}=0(1)$, 
\begin{equation}
\mathcal{N}_{\rm e/o}=\sum\limits_{k=0}^{\left\lfloor \frac{N}{2} \right\rfloor}{2(\Sites-M) \choose 2k+\eta_{{\rm e/o}}}{2M \choose N-2k-\eta_{{\rm e/o}}}
\label{eq:number_non-flipped_odd}
\end{equation}
gives the number of possible occupations with an even (e) and odd (o) number of particles in the $L-M$ states for which the spins are not flipped, and the echo probability, Eq.~(\ref{eq:def_mbse-probability}), for ${\bf X}={\bf T}$ takes the form \cite{suppl}
\begin{equation}
\mathcal{P}_{M,T}^{\rm echo}\left({\bf n};\tau\right)=1+\begin{cases}
\frac{\mathcal{N}_{\rm e}-\mathcal{N}_{\rm o}}{\mathcal{N}}, & \tau=0, \\
\frac{\left(-1\right)^{N}\mathcal{N}_{\rm e}^{(0)}}{\mathcal{N}}, & \tau\nsim0.
\end{cases}
\label{eq:echo-signal_tr}
\end{equation}

Finally, since contribution \ref{enum:si-pairing} requires ${\bf m}^\prime={\bf F}^{(M)}{\bf m}$ and ${\bf n}^\prime={\bf n}$, the corresponding sum of action differences can be evaluated in terms of the SOC phase $\varphi$ as
\begin{equation}
\Delta_{\gamma_{1}\gamma_{2}}^{\mathcal{T}\mathcal{S}_{\varphi}\gamma_{1}\mathcal{T}\mathcal{S}_{\varphi}\gamma_{2}}=2\varphi\sum\limits_{\sites=1}^{\Sites}\left[m_{\sites\downarrow}-\left({\bf F}^{(M)}{\bf m}\right)_{\sites\downarrow}\right].
\label{eq:action_difference_echo-si}
\end{equation}
As shown in \cite{suppl}, this contribution gives 
\begin{equation}
\mathcal{P}_{M,{\rm id}}^{\rm echo}\left({\bf n};\tau\right)=1+\begin{cases}
\frac{f(N,M;\varphi)}{\mathcal{N}}, & \tau=0 \\
\frac{\mathcal{N}_{\rm e}^{(0)}}{\mathcal{N}}, & \tau\nsim0,
\end{cases}
\label{eq:echo-signal_si}
\end{equation}
where the function 
\begin{equation}
\begin{split}
f&(N,M;\varphi)= \\
&\sum\limits_{k_{\uparrow}=0}^{N}\sum\limits_{k_{\downarrow}=0}^{N-k_{\uparrow}}{M \choose k_{\uparrow}}{M \choose k_{\downarrow}}{2L-2M \choose N-k_{\uparrow}-k_{\downarrow}}{\rm e}^{{\rm i}\varphi\left(k_{\downarrow}-k_{\uparrow}\right)}
\end{split}
\label{eq:echo-signal_si2}
\end{equation}
is given by a sum over the number $k_{\uparrow (\downarrow)}$ of spin-up(-down) particles in the flipped states. We note that the particle-hole symmetry of our results is guaranteed by the invariance of $\mathcal{P}_{M,X}^{\rm echo}$ under the replacement of the number of particles $N$ by the number of holes $2L-N$.

Our Eqs.~(\ref{eq:echo-signal_tr}) and (\ref{eq:echo-signal_si}) imply that the probabilities to measure the initial state or its spin-flipped counterpart display in most cases a peak or a dip well localized around $\tau=0$, the MBSE, and together with Eq.~(\ref{eq:echo-signal_si2}) constitute the main result of this paper. 

For the comparison of Eqs.~(\ref{eq:echo-signal_tr}) and (\ref{eq:echo-signal_si}) against numerical simulations in Fig.~\ref{fig:fig2}, the peak/background ratio is calculated as $\mathcal{P}^{\rm echo}(\tau=0)/\mathcal{P}^{\rm echo}(\tau\nsim0)$. In particular, for the case $M=L,\varphi=0$ shown in Fig.~\ref{fig:fig2}(a) the evaluation of $\mathcal{N}_{\rm e}^{(0)}$, $\mathcal{N}_{\rm e}$ and $\mathcal{N}_{\rm o}$ yields,
\begin{equation}
\label{eq:peak}
\left.\mathcal{P}_{L,X={\rm id},T}^{\rm echo}\left({\bf n},\tau\right)\right|_{\varphi=0}=\begin{cases}
2,& \tau=0 \\
1,& \tau\nsim0\quad\text{(N\text{ odd})}  \\
{L \choose \frac{N}{2}}/{2L \choose N},& \tau\nsim0\quad\text{(N\text{ even})},
\end{cases}
\end{equation}
in line with the results of the numerical simulation. The highly non-trivial (and universal) dependence of $\mathcal{P}_{M,{\rm id}}^{\rm echo}$ with $\varphi$ that follows from Eqs.~(\ref{eq:echo-signal_si}, \ref{eq:echo-signal_si2}) is depicted in Fig.~\ref{fig:fig2}b) for selected values $M=1,4,8$, and shows a remarkable agreement against the numerical simulations. This agreement is also seen in the detailed dependence of the echo peak on the number of sites $M$ for ${\bf X}={\rm id}$, Fig.~\ref{fig:fig2}c), and ${\bf X}={\bf T}$, Fig.~\ref{fig:fig2}d), for $\varphi=\pi/4$ \footnote{The case $
\left.\mathcal{P}_{L,{\rm id}}^{\rm echo}\left({\bf n},\tau=0\right)\right|_{\varphi=0 (\frac{\pi}{2})}=2(0)$ is due to coherent backscattering (Kramers degeneracy) characteristic of the Orthogonal (Symplectic) ensemble \cite{fermionic_propagator}.}. 
\begin{figure}[bbb]
\includegraphics[width=\columnwidth,angle=0]{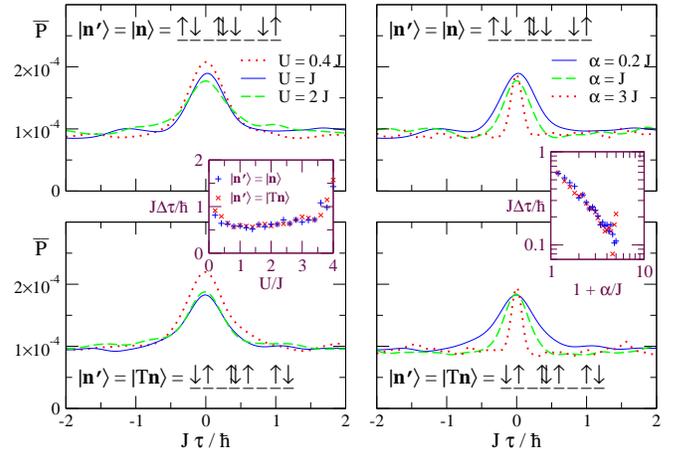}
\caption{Echo peak profiles, defined as the $\tau$-dependence of $\overline{P}$, for ${\bf X}={\rm id}$ (upper panels) and ${\bf X}={\bf T}$ (lower panels) for various values of the
interaction strength $U$ at $\alpha = J/5$ (left) and of the spin-orbit 
coupling $\alpha$ at $U = J$ (right). The insets show the fitted peak widths $\Delta \tau $\cite{captionfootnote2} for ${\bf X}={\rm id}$ ("$+$" symbols) and ${\bf X}={\bf T}$ ("$\times$" symbols) as a function of $U$ at $\alpha = J/5$ (left inset) and as a function
of $\alpha$ at $U = J$ (right inset).
While the peak width is fairly independent of $U$ within the wide parameter
range where the mean-field dynamics is expected to be chaotic, it is found to scale as $\tau\sim (J + \alpha)^{-1}$ 
[$\simeq 1/\mathrm{max}(J,\alpha)$ if $J \ll \alpha$ or $J \gg \alpha$]
with $J$ and $\alpha$, in agreement with the semiclassical prediction from Eq.~(\ref{eq:JD1}).}
\label{fig:widths}
\end{figure}

Finally, in order to estimate the $\tau$-dependence of the echo signals we expand the actions in Eq.~(\ref{eq:mbse_probability_sc_full}) to first order in time around $\tau=0$ and use the relation between the action and the conserved energy along classical trajectories $\partial R_{\gamma}(t)/\partial t=-H_{\rm MF}\left({\boldsymbol\psi}_{\gamma}^{\ast}(t),{\boldsymbol\psi}_{\gamma}(t)\right):=-E({\bf I}_{\gamma},{\boldsymbol \theta}_{\gamma})$. Using standard ergodic methods we obtain \cite{suppl}
\begin{equation}
\frac{P^{\rm (ii,iii)}\left({\bf n}',{\bf n};\tau\right)}{P^{{\rm (ii,iii)}}\left({\bf n}',{\bf n};0\right)}=\left|\int_{0}^{2\pi}{\rm e}^{\frac{i}{\hbar}E\left({\bf n}',{\boldsymbol \theta}\right)\tau}\frac{d^{2L}{\boldsymbol \theta}}{(2\pi)^{2L}}\right|^{2}
\label{eq:JD1}
\end{equation}
for the coherent contributions \ref{enum:tr-pairing} and \ref{enum:si-pairing}. Although this result suffers from an ambiguity \footnote{This ambiguity does not affect the calculation of the peak heights that depend only on the $\int_{0}^{t}{\rm d}s{\hbar\boldsymbol\theta}\cdot\dot{\bf I}$ part of $R_{\gamma}$} in the definition of the mean-field Hamiltonian $H_{\rm MF}$ \cite{fermionic_propagator}, two generic features of the MBSE width $\Delta \tau$ that follow from Eq.~(\ref{eq:JD1}) for the Hamiltonian in Eq.~(\ref{eq:quantum-hamiltonian}) are that it decays roughly as $1/{\rm Max}(\alpha,J)$ and that it is independent of the interaction strength $U$ \cite{suppl}. As shown in Fig.~\ref{fig:widths}, the numerical simulations show indeed these two features, confirming their universality as predicted by the semiclasical approach.     

%

In conclusion, we have predicted the existence of a quantum coherent effect that lifts the Hahn echo into the realm of interacting quantum systems. The many-body spin echo is a collective effect observable at the level of many-body dynamics where, due to quantum interference, the system echoes either its initial or its spin-flipped state after a sudden flip of the spins. Using a semiclassical approach based on interfering paths in Fock space, we show the relation between the many-body spin echo and anti-unitary symmetries, and predict that its signal has a universal dependence on few microscopic parameters if the classical mean field dynamics display chaotic behavior. This non-perturbative, chaotic regime where interactions, hopping and spin-orbit coupling are of similar strength is within reach of experimental realization using fermionic cold atoms. As all our analytical results show perfect agreement against extensive numerical simulations, the many-body spin echo offers the possibility to quantify many-body coherence in systems modeled by Fermi-Hubbard Hamiltonians, while establishing the long-sought connection between chaotic mean-field dynamics and universal coherent effects for fermionic fields.

%

\acknowledgements
We acknowledge support from DFG through SFB 689, and illuminating discussions with Harold Baranger.

\bibliographystyle{apsrev4-1}
\bibliography{mbse}

\begin{thebibliography}{42}%
\makeatletter
\providecommand \@ifxundefined [1]{%
 \@ifx{#1\undefined}
}%
\providecommand \@ifnum [1]{%
 \ifnum #1\expandafter \@firstoftwo
 \else \expandafter \@secondoftwo
 \fi
}%
\providecommand \@ifx [1]{%
 \ifx #1\expandafter \@firstoftwo
 \else \expandafter \@secondoftwo
 \fi
}%
\providecommand \natexlab [1]{#1}%
\providecommand \enquote  [1]{``#1''}%
\providecommand \bibnamefont  [1]{#1}%
\providecommand \bibfnamefont [1]{#1}%
\providecommand \citenamefont [1]{#1}%
\providecommand \href@noop [0]{\@secondoftwo}%
\providecommand \href [0]{\begingroup \@sanitize@url \@href}%
\providecommand \@href[1]{\@@startlink{#1}\@@href}%
\providecommand \@@href[1]{\endgroup#1\@@endlink}%
\providecommand \@sanitize@url [0]{\catcode `\\12\catcode `\$12\catcode
  `\&12\catcode `\#12\catcode `\^12\catcode `\_12\catcode `\%12\relax}%
\providecommand \@@startlink[1]{}%
\providecommand \@@endlink[0]{}%
\providecommand \url  [0]{\begingroup\@sanitize@url \@url }%
\providecommand \@url [1]{\endgroup\@href {#1}{\urlprefix }}%
\providecommand \urlprefix  [0]{URL }%
\providecommand \Eprint [0]{\href }%
\providecommand \doibase [0]{http://dx.doi.org/}%
\providecommand \selectlanguage [0]{\@gobble}%
\providecommand \bibinfo  [0]{\@secondoftwo}%
\providecommand \bibfield  [0]{\@secondoftwo}%
\providecommand \translation [1]{[#1]}%
\providecommand \BibitemOpen [0]{}%
\providecommand \bibitemStop [0]{}%
\providecommand \bibitemNoStop [0]{.\EOS\space}%
\providecommand \EOS [0]{\spacefactor3000\relax}%
\providecommand \BibitemShut  [1]{\csname bibitem#1\endcsname}%
\let\auto@bib@innerbib\@empty
\bibitem [{\citenamefont {Hahn}(1950)}]{Hahn-echo}%
  \BibitemOpen
  \bibfield  {author} {\bibinfo {author} {\bibfnamefont {E.~L.}\ \bibnamefont
  {Hahn}},\ }\href {\doibase 10.1103/PhysRev.80.580} {\bibfield  {journal}
  {\bibinfo  {journal} {Phys. Rev.}\ }\textbf {\bibinfo {volume} {80}},\
  \bibinfo {pages} {580} (\bibinfo {year} {1950})}\BibitemShut {NoStop}%
\bibitem [{\citenamefont {Prigodin}\ \emph {et~al.}(1994)\citenamefont
  {Prigodin}, \citenamefont {Altshuler}, \citenamefont {Efetov},\ and\
  \citenamefont {Iida}}]{mes-echo}%
  \BibitemOpen
  \bibfield  {author} {\bibinfo {author} {\bibfnamefont {V.~N.}\ \bibnamefont
  {Prigodin}}, \bibinfo {author} {\bibfnamefont {B.~L.}\ \bibnamefont
  {Altshuler}}, \bibinfo {author} {\bibfnamefont {K.~B.}\ \bibnamefont
  {Efetov}}, \ and\ \bibinfo {author} {\bibfnamefont {S.}~\bibnamefont
  {Iida}},\ }\href {\doibase 10.1103/PhysRevLett.72.546} {\bibfield  {journal}
  {\bibinfo  {journal} {Phys. Rev. Lett.}\ }\textbf {\bibinfo {volume} {72}},\
  \bibinfo {pages} {546} (\bibinfo {year} {1994})}\BibitemShut {NoStop}%
\bibitem [{\citenamefont {Peres}(1984)}]{loschmidt_echo_introduction}%
  \BibitemOpen
  \bibfield  {author} {\bibinfo {author} {\bibfnamefont {A.}~\bibnamefont
  {Peres}},\ }\href {\doibase 10.1103/PhysRevA.30.1610} {\bibfield  {journal}
  {\bibinfo  {journal} {Phys. Rev. A}\ }\textbf {\bibinfo {volume} {30}},\
  \bibinfo {pages} {1610} (\bibinfo {year} {1984})}\BibitemShut {NoStop}%
\bibitem [{\citenamefont {Malmberg}\ \emph {et~al.}(1968)\citenamefont
  {Malmberg}, \citenamefont {Wharton}, \citenamefont {Gould},\ and\
  \citenamefont {O'Neil}}]{plasmaecho}%
  \BibitemOpen
  \bibfield  {author} {\bibinfo {author} {\bibfnamefont {J.~H.}\ \bibnamefont
  {Malmberg}}, \bibinfo {author} {\bibfnamefont {C.~B.}\ \bibnamefont
  {Wharton}}, \bibinfo {author} {\bibfnamefont {R.~W.}\ \bibnamefont {Gould}},
  \ and\ \bibinfo {author} {\bibfnamefont {T.~M.}\ \bibnamefont {O'Neil}},\
  }\href {\doibase 10.1103/PhysRevLett.20.95} {\bibfield  {journal} {\bibinfo
  {journal} {Phys. Rev. Lett.}\ }\textbf {\bibinfo {volume} {20}},\ \bibinfo
  {pages} {95} (\bibinfo {year} {1968})}\BibitemShut {NoStop}%
\bibitem [{\citenamefont {Pierrat}\ \emph {et~al.}(2013)\citenamefont
  {Pierrat}, \citenamefont {Vandenbem}, \citenamefont {Fink},\ and\
  \citenamefont {Carminati}}]{Fink}%
  \BibitemOpen
  \bibfield  {author} {\bibinfo {author} {\bibfnamefont {R.}~\bibnamefont
  {Pierrat}}, \bibinfo {author} {\bibfnamefont {C.}~\bibnamefont {Vandenbem}},
  \bibinfo {author} {\bibfnamefont {M.}~\bibnamefont {Fink}}, \ and\ \bibinfo
  {author} {\bibfnamefont {R.}~\bibnamefont {Carminati}},\ }\href {\doibase
  10.1103/PhysRevA.87.041801} {\bibfield  {journal} {\bibinfo  {journal} {Phys.
  Rev. A}\ }\textbf {\bibinfo {volume} {87}},\ \bibinfo {pages} {041801}
  (\bibinfo {year} {2013})}\BibitemShut {NoStop}%
\bibitem [{\citenamefont {Klauder}\ and\ \citenamefont
  {Anderson}(1962)}]{KlauderAnderson}%
  \BibitemOpen
  \bibfield  {author} {\bibinfo {author} {\bibfnamefont {J.~R.}\ \bibnamefont
  {Klauder}}\ and\ \bibinfo {author} {\bibfnamefont {P.~W.}\ \bibnamefont
  {Anderson}},\ }\href {\doibase 10.1103/PhysRev.125.912} {\bibfield  {journal}
  {\bibinfo  {journal} {Phys. Rev.}\ }\textbf {\bibinfo {volume} {125}},\
  \bibinfo {pages} {912} (\bibinfo {year} {1962})}\BibitemShut {NoStop}%
\bibitem [{\citenamefont {Jalabert}\ and\ \citenamefont
  {Pastawski}(2001)}]{loschmidt_rodolfo}%
  \BibitemOpen
  \bibfield  {author} {\bibinfo {author} {\bibfnamefont {R.~A.}\ \bibnamefont
  {Jalabert}}\ and\ \bibinfo {author} {\bibfnamefont {H.~M.}\ \bibnamefont
  {Pastawski}},\ }\href {\doibase 10.1103/PhysRevLett.86.2490} {\bibfield
  {journal} {\bibinfo  {journal} {Phys. Rev. Lett.}\ }\textbf {\bibinfo
  {volume} {86}},\ \bibinfo {pages} {2490} (\bibinfo {year}
  {2001})}\BibitemShut {NoStop}%
\bibitem [{\citenamefont {Abe}\ \emph {et~al.}(2004)\citenamefont {Abe},
  \citenamefont {Itoh}, \citenamefont {Isoya},\ and\ \citenamefont
  {Yamasaki}}]{hahn-decay}%
  \BibitemOpen
  \bibfield  {author} {\bibinfo {author} {\bibfnamefont {E.}~\bibnamefont
  {Abe}}, \bibinfo {author} {\bibfnamefont {K.~M.}\ \bibnamefont {Itoh}},
  \bibinfo {author} {\bibfnamefont {J.}~\bibnamefont {Isoya}}, \ and\ \bibinfo
  {author} {\bibfnamefont {S.}~\bibnamefont {Yamasaki}},\ }\href {\doibase
  10.1103/PhysRevB.70.033204} {\bibfield  {journal} {\bibinfo  {journal} {Phys.
  Rev. B}\ }\textbf {\bibinfo {volume} {70}},\ \bibinfo {pages} {033204}
  (\bibinfo {year} {2004})}\BibitemShut {NoStop}%
\bibitem [{\citenamefont {Witzel}\ \emph {et~al.}(2005)\citenamefont {Witzel},
  \citenamefont {de~Sousa},\ and\ \citenamefont {{Das
  Sarma}}}]{decoherence_hahn2}%
  \BibitemOpen
  \bibfield  {author} {\bibinfo {author} {\bibfnamefont {W.~M.}\ \bibnamefont
  {Witzel}}, \bibinfo {author} {\bibfnamefont {R.}~\bibnamefont {de~Sousa}}, \
  and\ \bibinfo {author} {\bibfnamefont {S.}~\bibnamefont {{Das Sarma}}},\
  }\href {\doibase 10.1103/PhysRevB.72.161306} {\bibfield  {journal} {\bibinfo
  {journal} {Phys. Rev. B}\ }\textbf {\bibinfo {volume} {72}},\ \bibinfo
  {pages} {161306} (\bibinfo {year} {2005})}\BibitemShut {NoStop}%
\bibitem [{\citenamefont {Micklitz}\ \emph {et~al.}(2015)\citenamefont
  {Micklitz}, \citenamefont {M\"uller},\ and\ \citenamefont {Altland}}]{Cord}%
  \BibitemOpen
  \bibfield  {author} {\bibinfo {author} {\bibfnamefont {T.}~\bibnamefont
  {Micklitz}}, \bibinfo {author} {\bibfnamefont {C.~A.}\ \bibnamefont
  {M\"uller}}, \ and\ \bibinfo {author} {\bibfnamefont {A.}~\bibnamefont
  {Altland}},\ }\href {\doibase 10.1103/PhysRevB.91.064203} {\bibfield
  {journal} {\bibinfo  {journal} {Phys. Rev. B}\ }\textbf {\bibinfo {volume}
  {91}},\ \bibinfo {pages} {064203} (\bibinfo {year} {2015})}\BibitemShut
  {NoStop}%
\bibitem [{\citenamefont {M\"uller}\ \emph {et~al.}(2015)\citenamefont
  {M\"uller}, \citenamefont {Richard}, \citenamefont {Volchkov}, \citenamefont
  {Denechaud}, \citenamefont {Bouyer}, \citenamefont {Aspect},\ and\
  \citenamefont {Josse}}]{Cord_exp}%
  \BibitemOpen
  \bibfield  {author} {\bibinfo {author} {\bibfnamefont {K.}~\bibnamefont
  {M\"uller}}, \bibinfo {author} {\bibfnamefont {J.}~\bibnamefont {Richard}},
  \bibinfo {author} {\bibfnamefont {V.~V.}\ \bibnamefont {Volchkov}}, \bibinfo
  {author} {\bibfnamefont {V.}~\bibnamefont {Denechaud}}, \bibinfo {author}
  {\bibfnamefont {P.}~\bibnamefont {Bouyer}}, \bibinfo {author} {\bibfnamefont
  {A.}~\bibnamefont {Aspect}}, \ and\ \bibinfo {author} {\bibfnamefont
  {V.}~\bibnamefont {Josse}},\ }\href {\doibase 10.1103/PhysRevLett.114.205301}
  {\bibfield  {journal} {\bibinfo  {journal} {Phys. Rev. Lett.}\ }\textbf
  {\bibinfo {volume} {114}},\ \bibinfo {pages} {205301} (\bibinfo {year}
  {2015})}\BibitemShut {NoStop}%
\bibitem [{\citenamefont {Yi}\ \emph {et~al.}(2007)\citenamefont {Yi},
  \citenamefont {Wang},\ and\ \citenamefont {Wang}}]{coupled_hahn_echo}%
  \BibitemOpen
  \bibfield  {author} {\bibinfo {author} {\bibfnamefont {X.~X.}\ \bibnamefont
  {Yi}}, \bibinfo {author} {\bibfnamefont {H.}~\bibnamefont {Wang}}, \ and\
  \bibinfo {author} {\bibfnamefont {W.}~\bibnamefont {Wang}},\ }\href {\doibase
  10.1140/epjd/e2007-00266-6} {\bibfield  {journal} {\bibinfo  {journal} {The
  European Physical Journal D}\ }\textbf {\bibinfo {volume} {45}},\ \bibinfo
  {pages} {355} (\bibinfo {year} {2007})}\BibitemShut {NoStop}%
\bibitem [{\citenamefont {Ma}\ \emph {et~al.}(2014)\citenamefont {Ma},
  \citenamefont {Wolfowicz}, \citenamefont {Zhao}, \citenamefont {Li},
  \citenamefont {Morton},\ and\ \citenamefont {Liu}}]{spin-decoherence}%
  \BibitemOpen
  \bibfield  {author} {\bibinfo {author} {\bibfnamefont {W.-L.}\ \bibnamefont
  {Ma}}, \bibinfo {author} {\bibfnamefont {G.}~\bibnamefont {Wolfowicz}},
  \bibinfo {author} {\bibfnamefont {N.}~\bibnamefont {Zhao}}, \bibinfo {author}
  {\bibfnamefont {S.-S.}\ \bibnamefont {Li}}, \bibinfo {author} {\bibfnamefont
  {J.~J.~L.}\ \bibnamefont {Morton}}, \ and\ \bibinfo {author} {\bibfnamefont
  {R.-B.}\ \bibnamefont {Liu}},\ }\href {\doibase 10.1038/ncomms5822}
  {\bibfield  {journal} {\bibinfo  {journal} {{Nat.~Commun}}\ }\textbf
  {\bibinfo {volume} {{5}}},\ \bibinfo {pages} {{4822}} (\bibinfo {year}
  {{2014}})}\BibitemShut {NoStop}%
\bibitem [{\citenamefont {Luan}\ \emph {et~al.}(2015)\citenamefont {Luan},
  \citenamefont {Grinolds}, \citenamefont {Hong}, \citenamefont {Maletinsky},
  \citenamefont {Walsworth},\ and\ \citenamefont {Yacoby}}]{spins_diamond}%
  \BibitemOpen
  \bibfield  {author} {\bibinfo {author} {\bibfnamefont {L.}~\bibnamefont
  {Luan}}, \bibinfo {author} {\bibfnamefont {M.~S.}\ \bibnamefont {Grinolds}},
  \bibinfo {author} {\bibfnamefont {S.}~\bibnamefont {Hong}}, \bibinfo {author}
  {\bibfnamefont {P.}~\bibnamefont {Maletinsky}}, \bibinfo {author}
  {\bibfnamefont {R.~L.}\ \bibnamefont {Walsworth}}, \ and\ \bibinfo {author}
  {\bibfnamefont {A.}~\bibnamefont {Yacoby}},\ }\href {\doibase
  10.1038/srep08119} {\bibfield  {journal} {\bibinfo  {journal} {{Sci.~Rep.}}\
  }\textbf {\bibinfo {volume} {{5}}},\ \bibinfo {pages} {{8119 }} (\bibinfo
  {year} {{2015}})}\BibitemShut {NoStop}%
\bibitem [{\citenamefont {{Solaro}}\ \emph {et~al.}(2016)\citenamefont
  {{Solaro}}, \citenamefont {{Bonnin}}, \citenamefont {{Combes}}, \citenamefont
  {{Lopez}}, \citenamefont {{Alauze}}, \citenamefont {{Fuchs}}, \citenamefont
  {{Pi{\'e}chon}},\ and\ \citenamefont {{Pereira dos
  Santos}}}]{hahn-echo_trapped_atoms}%
  \BibitemOpen
  \bibfield  {author} {\bibinfo {author} {\bibfnamefont {C.}~\bibnamefont
  {{Solaro}}}, \bibinfo {author} {\bibfnamefont {A.}~\bibnamefont {{Bonnin}}},
  \bibinfo {author} {\bibfnamefont {F.}~\bibnamefont {{Combes}}}, \bibinfo
  {author} {\bibfnamefont {M.}~\bibnamefont {{Lopez}}}, \bibinfo {author}
  {\bibfnamefont {X.}~\bibnamefont {{Alauze}}}, \bibinfo {author}
  {\bibfnamefont {J.-N.}\ \bibnamefont {{Fuchs}}}, \bibinfo {author}
  {\bibfnamefont {F.}~\bibnamefont {{Pi{\'e}chon}}}, \ and\ \bibinfo {author}
  {\bibfnamefont {F.}~\bibnamefont {{Pereira dos Santos}}},\ }\href@noop {}
  {\enquote {\bibinfo {title} {{Competition between Spin Echo and Spin
  Self-Rephasing in a Trapped Atom Interferometer}},}\ } (\bibinfo {year}
  {2016}),\ \Eprint {http://arxiv.org/abs/1606.00218} {arXiv:1606.00218
  [physics.atom-ph]} \BibitemShut {NoStop}%
\bibitem [{\citenamefont {Knap}\ \emph {et~al.}(2013)\citenamefont {Knap},
  \citenamefont {Kantian}, \citenamefont {Giamarchi}, \citenamefont {Bloch},
  \citenamefont {Lukin},\ and\ \citenamefont
  {Demler}}]{Knap_selfadvertisement1}%
  \BibitemOpen
  \bibfield  {author} {\bibinfo {author} {\bibfnamefont {M.}~\bibnamefont
  {Knap}}, \bibinfo {author} {\bibfnamefont {A.}~\bibnamefont {Kantian}},
  \bibinfo {author} {\bibfnamefont {T.}~\bibnamefont {Giamarchi}}, \bibinfo
  {author} {\bibfnamefont {I.}~\bibnamefont {Bloch}}, \bibinfo {author}
  {\bibfnamefont {M.~D.}\ \bibnamefont {Lukin}}, \ and\ \bibinfo {author}
  {\bibfnamefont {E.}~\bibnamefont {Demler}},\ }\href {\doibase
  10.1103/PhysRevLett.111.147205} {\bibfield  {journal} {\bibinfo  {journal}
  {Phys. Rev. Lett.}\ }\textbf {\bibinfo {volume} {111}},\ \bibinfo {pages}
  {147205} (\bibinfo {year} {2013})}\BibitemShut {NoStop}%
\bibitem [{\citenamefont {Serbyn}\ \emph {et~al.}(2014)\citenamefont {Serbyn},
  \citenamefont {Knap}, \citenamefont {Gopalakrishnan}, \citenamefont
  {Papi\ifmmode~\acute{c}\else \'{c}\fi{}}, \citenamefont {Yao}, \citenamefont
  {Laumann}, \citenamefont {Abanin}, \citenamefont {Lukin},\ and\ \citenamefont
  {Demler}}]{Knap_selfadvertisement2}%
  \BibitemOpen
  \bibfield  {author} {\bibinfo {author} {\bibfnamefont {M.}~\bibnamefont
  {Serbyn}}, \bibinfo {author} {\bibfnamefont {M.}~\bibnamefont {Knap}},
  \bibinfo {author} {\bibfnamefont {S.}~\bibnamefont {Gopalakrishnan}},
  \bibinfo {author} {\bibfnamefont {Z.}~\bibnamefont
  {Papi\ifmmode~\acute{c}\else \'{c}\fi{}}}, \bibinfo {author} {\bibfnamefont
  {N.~Y.}\ \bibnamefont {Yao}}, \bibinfo {author} {\bibfnamefont {C.~R.}\
  \bibnamefont {Laumann}}, \bibinfo {author} {\bibfnamefont {D.~A.}\
  \bibnamefont {Abanin}}, \bibinfo {author} {\bibfnamefont {M.~D.}\
  \bibnamefont {Lukin}}, \ and\ \bibinfo {author} {\bibfnamefont {E.~A.}\
  \bibnamefont {Demler}},\ }\href {\doibase 10.1103/PhysRevLett.113.147204}
  {\bibfield  {journal} {\bibinfo  {journal} {Phys. Rev. Lett.}\ }\textbf
  {\bibinfo {volume} {113}},\ \bibinfo {pages} {147204} (\bibinfo {year}
  {2014})}\BibitemShut {NoStop}%
\bibitem [{\citenamefont {Bychkov}\ and\ \citenamefont
  {Rashba}(1984)}]{Rashba}%
  \BibitemOpen
  \bibfield  {author} {\bibinfo {author} {\bibfnamefont {Y.~A.}\ \bibnamefont
  {Bychkov}}\ and\ \bibinfo {author} {\bibfnamefont {E.~I.}\ \bibnamefont
  {Rashba}},\ }\href@noop {} {\bibfield  {journal} {\bibinfo  {journal}
  {J.~Phys.~C: Solid State Phys.}\ }\textbf {\bibinfo {volume} {17}},\ \bibinfo
  {pages} {6039} (\bibinfo {year} {1984})}\BibitemShut {NoStop}%
\bibitem [{\citenamefont {Dresselhaus}(1955)}]{Dresselhaus}%
  \BibitemOpen
  \bibfield  {author} {\bibinfo {author} {\bibfnamefont {G.}~\bibnamefont
  {Dresselhaus}},\ }\href {\doibase 10.1103/PhysRev.100.580} {\bibfield
  {journal} {\bibinfo  {journal} {Phys. Rev.}\ }\textbf {\bibinfo {volume}
  {100}},\ \bibinfo {pages} {580} (\bibinfo {year} {1955})}\BibitemShut
  {NoStop}%
\bibitem [{\citenamefont {Behbood}\ \emph {et~al.}(2014)\citenamefont
  {Behbood}, \citenamefont {{Martin Ciurana}}, \citenamefont {Colangelo},
  \citenamefont {Napolitano}, \citenamefont {T{\'o}th}, \citenamefont
  {Sewell},\ and\ \citenamefont {Mitchell}}]{macroscopic_singlet}%
  \BibitemOpen
  \bibfield  {author} {\bibinfo {author} {\bibfnamefont {N.}~\bibnamefont
  {Behbood}}, \bibinfo {author} {\bibfnamefont {F.}~\bibnamefont {{Martin
  Ciurana}}}, \bibinfo {author} {\bibfnamefont {G.}~\bibnamefont {Colangelo}},
  \bibinfo {author} {\bibfnamefont {M.}~\bibnamefont {Napolitano}}, \bibinfo
  {author} {\bibfnamefont {G.}~\bibnamefont {T{\'o}th}}, \bibinfo {author}
  {\bibfnamefont {R.~J.}\ \bibnamefont {Sewell}}, \ and\ \bibinfo {author}
  {\bibfnamefont {M.~W.}\ \bibnamefont {Mitchell}},\ }\href {\doibase
  10.1103/PhysRevLett.113.093601} {\bibfield  {journal} {\bibinfo  {journal}
  {Phys. Rev. Lett.}\ }\textbf {\bibinfo {volume} {113}},\ \bibinfo {pages}
  {093601} (\bibinfo {year} {2014})}\BibitemShut {NoStop}%
\bibitem [{\citenamefont {Lin}\ \emph {et~al.}(2011)\citenamefont {Lin},
  \citenamefont {Jimenez-Garcia},\ and\ \citenamefont
  {Spielman}}]{soi_ultracold_atoms}%
  \BibitemOpen
  \bibfield  {author} {\bibinfo {author} {\bibfnamefont {Y.-J.}\ \bibnamefont
  {Lin}}, \bibinfo {author} {\bibfnamefont {K.}~\bibnamefont {Jimenez-Garcia}},
  \ and\ \bibinfo {author} {\bibfnamefont {I.~B.}\ \bibnamefont {Spielman}},\
  }\href {\doibase 10.1038/nature09887} {\bibfield  {journal} {\bibinfo
  {journal} {Nature}\ }\textbf {\bibinfo {volume} {471}},\ \bibinfo {pages}
  {83} (\bibinfo {year} {2011})}\BibitemShut {NoStop}%
\bibitem [{\citenamefont {Sau}\ \emph {et~al.}(2011)\citenamefont {Sau},
  \citenamefont {Sensarma}, \citenamefont {Powell}, \citenamefont {Spielman},\
  and\ \citenamefont {{Das Sarma}}}]{soi_ultracold_atoms2}%
  \BibitemOpen
  \bibfield  {author} {\bibinfo {author} {\bibfnamefont {J.~D.}\ \bibnamefont
  {Sau}}, \bibinfo {author} {\bibfnamefont {R.}~\bibnamefont {Sensarma}},
  \bibinfo {author} {\bibfnamefont {S.}~\bibnamefont {Powell}}, \bibinfo
  {author} {\bibfnamefont {I.~B.}\ \bibnamefont {Spielman}}, \ and\ \bibinfo
  {author} {\bibfnamefont {S.}~\bibnamefont {{Das Sarma}}},\ }\href {\doibase
  10.1103/PhysRevB.83.140510} {\bibfield  {journal} {\bibinfo  {journal} {Phys.
  Rev. B}\ }\textbf {\bibinfo {volume} {83}},\ \bibinfo {pages} {140510}
  (\bibinfo {year} {2011})}\BibitemShut {NoStop}%
\bibitem [{\citenamefont {Campbell}\ \emph {et~al.}(2011)\citenamefont
  {Campbell}, \citenamefont {{Juzeli\ifmmode \bar{u}\else \={u}\fi{}nas}},\
  and\ \citenamefont {Spielman}}]{soi_ultracold_atoms3}%
  \BibitemOpen
  \bibfield  {author} {\bibinfo {author} {\bibfnamefont {D.~L.}\ \bibnamefont
  {Campbell}}, \bibinfo {author} {\bibfnamefont {G.}~\bibnamefont
  {{Juzeli\ifmmode \bar{u}\else \={u}\fi{}nas}}}, \ and\ \bibinfo {author}
  {\bibfnamefont {I.~B.}\ \bibnamefont {Spielman}},\ }\href {\doibase
  10.1103/PhysRevA.84.025602} {\bibfield  {journal} {\bibinfo  {journal} {Phys.
  Rev. A}\ }\textbf {\bibinfo {volume} {84}},\ \bibinfo {pages} {025602}
  (\bibinfo {year} {2011})}\BibitemShut {NoStop}%
\bibitem [{\citenamefont {Lewenstein}\ \emph {et~al.}(2012)\citenamefont
  {Lewenstein}, \citenamefont {{Sanpera Trigueros}},\ and\ \citenamefont
  {Ahufinger}}]{Lewenstein:2012}%
  \BibitemOpen
  \bibfield  {author} {\bibinfo {author} {\bibfnamefont {M.}~\bibnamefont
  {Lewenstein}}, \bibinfo {author} {\bibfnamefont {A.}~\bibnamefont {{Sanpera
  Trigueros}}}, \ and\ \bibinfo {author} {\bibfnamefont {V.}~\bibnamefont
  {Ahufinger}},\ }\href
  {http://bvbr.bib-bvb.de:8991/F?func=service&doc_library=BVB01&local_base=BVB01&doc_number=024894719&line_number=0001&func_code=DB_RECORDS&service_type=MEDIA}
  {\emph {\bibinfo {title} {Ultracold atoms in optical lattices: simulating
  quantum many-body systems}}},\ \bibinfo {edition} {1st}\ ed.\ (\bibinfo
  {publisher} {Oxford Univ. Press},\ \bibinfo {address} {Oxford},\ \bibinfo
  {year} {2012})\BibitemShut {NoStop}%
\bibitem [{\citenamefont {Lewenstein}\ \emph {et~al.}(2007)\citenamefont
  {Lewenstein}, \citenamefont {Sanpera}, \citenamefont {Ahufinger},
  \citenamefont {Damski}, \citenamefont {Sen(De)},\ and\ \citenamefont
  {Sen}}]{ultracold-atoms_mimicking_condensed_matter-review}%
  \BibitemOpen
  \bibfield  {author} {\bibinfo {author} {\bibfnamefont {M.}~\bibnamefont
  {Lewenstein}}, \bibinfo {author} {\bibfnamefont {A.}~\bibnamefont {Sanpera}},
  \bibinfo {author} {\bibfnamefont {V.}~\bibnamefont {Ahufinger}}, \bibinfo
  {author} {\bibfnamefont {B.}~\bibnamefont {Damski}}, \bibinfo {author}
  {\bibfnamefont {A.}~\bibnamefont {Sen(De)}}, \ and\ \bibinfo {author}
  {\bibfnamefont {U.}~\bibnamefont {Sen}},\ }\href@noop {} {\bibfield
  {journal} {\bibinfo  {journal} {Advances in Physics}\ }\textbf {\bibinfo
  {volume} {56}},\ \bibinfo {pages} {243} (\bibinfo {year} {2007})}\BibitemShut
  {NoStop}%
\bibitem [{Note1()}]{Note1}%
  \BibitemOpen
  \bibinfo {note} {The result will not depend on the choice of the sites. To
  simplify the notation we take the first $M$.}\BibitemShut {Stop}%
\bibitem [{Note2()}]{Note2}%
  \BibitemOpen
  \bibinfo {note} {Adding an additional phase factor ${\protect \rm
  e}^{{\protect \rm i}\chi }$ to the spin flip operator, $\protect \mathaccentV
  {hat}05E{A}_{l}=\left (1-\protect \mathaccentV {hat}05E{n}_{l,\delimiter
  "3222378 }^{}\right )\left (1-\protect \mathaccentV {hat}05E{n}_{l,\delimiter
  "3223379 }^{}\right )+\protect \mathaccentV {hat}05E{n}_{l,\delimiter
  "3222378 }^{}\protect \mathaccentV {hat}05E{n}_{l,\delimiter "3223379
  }^{}+{\protect \rm e}^{{\protect \rm i}\chi }\protect \mathaccentV
  {hat}05E{c}_{l,\delimiter "3222378 }^{\dagger }\protect \mathaccentV
  {hat}05E{c}_{l,\delimiter "3223379 }^{}+{\protect \rm e}^{-{\protect \rm
  i}\chi }\protect \mathaccentV {hat}05E{c}_{l,\delimiter "3223379 }^{\dagger
  }\protect \mathaccentV {hat}05E{c}_{l,\delimiter "3222378 }^{}$ corresponds
  to changing the Rashba phase $\varphi =\protect \qopname \relax o{arg}(\kappa
  _{\delimiter "3222378 \delimiter "3223379 })$ to $\varphi +\chi
  $.}\BibitemShut {Stop}%
\bibitem [{\citenamefont {Engl}\ \emph
  {et~al.}(2014{\natexlab{a}})\citenamefont {Engl}, \citenamefont
  {Pl{\"o}{\ss}l}, \citenamefont {Urbina},\ and\ \citenamefont
  {Richter}}]{fermionic_propagator}%
  \BibitemOpen
  \bibfield  {author} {\bibinfo {author} {\bibfnamefont {T.}~\bibnamefont
  {Engl}}, \bibinfo {author} {\bibfnamefont {P.}~\bibnamefont {Pl{\"o}{\ss}l}},
  \bibinfo {author} {\bibfnamefont {J.}~\bibnamefont {Urbina}}, \ and\ \bibinfo
  {author} {\bibfnamefont {K.}~\bibnamefont {Richter}},\ }\href
  {http://dx.doi.org/10.1007/s00214-014-1563-9} {\bibfield  {journal} {\bibinfo
   {journal} {Theoretical Chemistry Accounts}\ }\textbf {\bibinfo {volume}
  {133}},\ \bibinfo {eid} {1563} (\bibinfo {year}
  {2014}{\natexlab{a}})}\BibitemShut {NoStop}%
\bibitem [{\citenamefont {Grosse-Holz}\ \emph {et~al.}(2015)\citenamefont
  {Grosse-Holz}, \citenamefont {Engl}, \citenamefont {Richter},\ and\
  \citenamefont {Urbina}}]{simon}%
  \BibitemOpen
  \bibfield  {author} {\bibinfo {author} {\bibfnamefont {S.}~\bibnamefont
  {Grosse-Holz}}, \bibinfo {author} {\bibfnamefont {T.}~\bibnamefont {Engl}},
  \bibinfo {author} {\bibfnamefont {K.}~\bibnamefont {Richter}}, \ and\
  \bibinfo {author} {\bibfnamefont {J.~D.}\ \bibnamefont {Urbina}},\
  }\href@noop {} {\bibfield  {journal} {\bibinfo  {journal} {Acta Physica
  Polonica A}\ }\textbf {\bibinfo {volume} {128}},\ \bibinfo {pages} {994}
  (\bibinfo {year} {2015})}\BibitemShut {NoStop}%
\bibitem [{\citenamefont {Gutzwiller}(1990)}]{BGS}%
  \BibitemOpen
  \bibfield  {author} {\bibinfo {author} {\bibfnamefont {M.~C.}\ \bibnamefont
  {Gutzwiller}},\ }\href@noop {} {\emph {\bibinfo {title} {{Chaos in Classical
  and Quantum Mechanics}}}}\ (\bibinfo  {publisher} {Springer},\ \bibinfo
  {year} {1990})\BibitemShut {NoStop}%
\bibitem [{\citenamefont {Sieber}\ and\ \citenamefont
  {Richter}(2001)}]{SR-pairs}%
  \BibitemOpen
  \bibfield  {author} {\bibinfo {author} {\bibfnamefont {M.}~\bibnamefont
  {Sieber}}\ and\ \bibinfo {author} {\bibfnamefont {K.}~\bibnamefont
  {Richter}},\ }\href {http://stacks.iop.org/1402-4896/2001/i=T90/a=018}
  {\bibfield  {journal} {\bibinfo  {journal} {Physica Scripta}\ }\textbf
  {\bibinfo {volume} {T90}},\ \bibinfo {pages} {128} (\bibinfo {year}
  {2001})}\BibitemShut {NoStop}%
\bibitem [{\citenamefont {Richter}\ and\ \citenamefont
  {Sieber}(2002)}]{quantumchaostransportKlaus}%
  \BibitemOpen
  \bibfield  {author} {\bibinfo {author} {\bibfnamefont {K.}~\bibnamefont
  {Richter}}\ and\ \bibinfo {author} {\bibfnamefont {M.}~\bibnamefont
  {Sieber}},\ }\href@noop {} {\bibfield  {journal} {\bibinfo  {journal}
  {Phys.~Rev.~Lett.}\ }\textbf {\bibinfo {volume} {89}},\ \bibinfo {pages}
  {206801} (\bibinfo {year} {2002})}\BibitemShut {NoStop}%
\bibitem [{\citenamefont {Turek}\ \emph {et~al.}(2005)\citenamefont {Turek},
  \citenamefont {Spehner}, \citenamefont {M{\"u}ller},\ and\ \citenamefont
  {Richter}}]{SR-pairs_dg3}%
  \BibitemOpen
  \bibfield  {author} {\bibinfo {author} {\bibfnamefont {M.}~\bibnamefont
  {Turek}}, \bibinfo {author} {\bibfnamefont {D.}~\bibnamefont {Spehner}},
  \bibinfo {author} {\bibfnamefont {S.}~\bibnamefont {M{\"u}ller}}, \ and\
  \bibinfo {author} {\bibfnamefont {K.}~\bibnamefont {Richter}},\ }\href
  {\doibase 10.1103/PhysRevE.71.016210} {\bibfield  {journal} {\bibinfo
  {journal} {Phys. Rev. E}\ }\textbf {\bibinfo {volume} {71}},\ \bibinfo
  {pages} {016210} (\bibinfo {year} {2005})}\BibitemShut {NoStop}%
\bibitem [{\citenamefont {Heusler}\ \emph {et~al.}(2004)\citenamefont
  {Heusler}, \citenamefont {M{\"u}ller}, \citenamefont {Braun},\ and\
  \citenamefont {Haake}}]{3-encounter}%
  \BibitemOpen
  \bibfield  {author} {\bibinfo {author} {\bibfnamefont {S.}~\bibnamefont
  {Heusler}}, \bibinfo {author} {\bibfnamefont {S.}~\bibnamefont {M{\"u}ller}},
  \bibinfo {author} {\bibfnamefont {P.}~\bibnamefont {Braun}}, \ and\ \bibinfo
  {author} {\bibfnamefont {F.}~\bibnamefont {Haake}},\ }\href
  {http://stacks.iop.org/0305-4470/37/i=3/a=L02} {\bibfield  {journal}
  {\bibinfo  {journal} {J.~Phys.~A: Math.~Gen.}\ }\textbf {\bibinfo {volume}
  {37}},\ \bibinfo {pages} {L31} (\bibinfo {year} {2004})}\BibitemShut
  {NoStop}%
\bibitem [{\citenamefont {Heusler}\ \emph {et~al.}(2006)\citenamefont
  {Heusler}, \citenamefont {M{\"u}ller}, \citenamefont {Braun},\ and\
  \citenamefont {Haake}}]{loops_transport}%
  \BibitemOpen
  \bibfield  {author} {\bibinfo {author} {\bibfnamefont {S.}~\bibnamefont
  {Heusler}}, \bibinfo {author} {\bibfnamefont {S.}~\bibnamefont {M{\"u}ller}},
  \bibinfo {author} {\bibfnamefont {P.}~\bibnamefont {Braun}}, \ and\ \bibinfo
  {author} {\bibfnamefont {F.}~\bibnamefont {Haake}},\ }\href {\doibase
  10.1103/PhysRevLett.96.066804} {\bibfield  {journal} {\bibinfo  {journal}
  {Phys. Rev. Lett.}\ }\textbf {\bibinfo {volume} {96}},\ \bibinfo {pages}
  {066804} (\bibinfo {year} {2006})}\BibitemShut {NoStop}%
\bibitem [{\citenamefont {Berry}(1985)}]{da}%
  \BibitemOpen
  \bibfield  {author} {\bibinfo {author} {\bibfnamefont {M.~V.}\ \bibnamefont
  {Berry}},\ }\href
  {http://rspa.royalsocietypublishing.org/content/400/1819/229} {\bibfield
  {journal} {\bibinfo  {journal} {Proc. R. Soc. Lond. A}\ }\textbf {\bibinfo
  {volume} {400}},\ \bibinfo {pages} {229} (\bibinfo {year}
  {1985})}\BibitemShut {NoStop}%
\bibitem [{\citenamefont {Engl}\ \emph
  {et~al.}(2014{\natexlab{b}})\citenamefont {Engl}, \citenamefont {Dujardin},
  \citenamefont {Arg{\"u}elles}, \citenamefont {Schlagheck}, \citenamefont
  {Richter},\ and\ \citenamefont {Urbina}}]{cbs_fock}%
  \BibitemOpen
  \bibfield  {author} {\bibinfo {author} {\bibfnamefont {T.}~\bibnamefont
  {Engl}}, \bibinfo {author} {\bibfnamefont {J.}~\bibnamefont {Dujardin}},
  \bibinfo {author} {\bibfnamefont {A.}~\bibnamefont {Arg{\"u}elles}}, \bibinfo
  {author} {\bibfnamefont {P.}~\bibnamefont {Schlagheck}}, \bibinfo {author}
  {\bibfnamefont {K.}~\bibnamefont {Richter}}, \ and\ \bibinfo {author}
  {\bibfnamefont {J.~D.}\ \bibnamefont {Urbina}},\ }\href {\doibase
  10.1103/PhysRevLett.112.140403} {\bibfield  {journal} {\bibinfo  {journal}
  {Phys. Rev. Lett.}\ }\textbf {\bibinfo {volume} {112}},\ \bibinfo {pages}
  {140403} (\bibinfo {year} {2014}{\natexlab{b}})}\BibitemShut {NoStop}%
\bibitem [{Note3()}]{Note3}%
  \BibitemOpen
  \bibinfo {note} {For $M=0$ this yields the transition probability
  $P=p_{\protect \rm cl}(1+\delta _{{\protect \bf n}^{\prime },{\protect \bf
  n}}+\left (-1\right )^{N}\delta _{{\protect \bf n}^{\prime },{\protect \bf
  F}^{(L)}{\protect \bf n}})$ along the lines of the studies of coherent
  forward- and backscattering.~\cite
  {cbs_fock,fermionic_propagator}}\BibitemShut {NoStop}%
\bibitem [{sup()}]{suppl}%
  \BibitemOpen
  \href@noop {} {}\bibinfo {note} {Supplementary}\BibitemShut {NoStop}%
\bibitem [{Note4()}]{Note4}%
  \BibitemOpen
  \bibinfo {note} {The case $ \left .\protect \mathcal {P}_{L,{\protect \rm
  id}}^{\protect \rm echo}\left ({\protect \bf n},\tau =0\right )\right
  |_{\varphi =0 (\protect \frac {\pi }{2})}=2(0)$ is due to coherent
  backscattering (Kramers degeneracy) characteristic of the Orthogonal
  (Symplectic) ensemble \cite {fermionic_propagator}.}\BibitemShut {Stop}%
\bibitem [{cap()}]{captionfootnote2}%
  \BibitemOpen
  \href@noop {} {}\bibinfo {note} {{T}he peak profiles are fitted to a
  Lorentzian defined on top of a flat background, using the background level
  and the width of the Lorentzian as adjustable parameters (with the maximum
  relative enhancement being fixed to 2). While this fit works well for the
  peak profiles shown in Fig. 4, it becomes less convincing at large $\alpha/J
  \simeq 4$ where side peaks arise in the profile.}\BibitemShut {Stop}%
\bibitem [{Note5()}]{Note5}%
  \BibitemOpen
  \bibinfo {note} {This ambiguity does not affect the calculation of the peak
  heights that depend only on the $\DOTSI \intop \ilimits@ _{0}^{t}{\protect
  \rm d}s{\hbar \protect \boldsymbol \theta }\cdot \protect \mathaccentV
  {dot}05F{\protect \bf I}$ part of $R_{\gamma }$}\BibitemShut {NoStop}%
\end{thebibliography}%
\end{document}


\renewcommand{\theequation}{S\arabic{equation}}
\title{Many-Body Spin Echo: \\ Supplemental material}
\newcommand{\RegensburgUniversity}{Institut f\"ur Theoretische Physik, 
Universit\"at Regensburg, D-93040 Regensburg, Germany}
\newcommand{\LiegeUniversity}{Institute of Theoretical Physics, Liege University, Liege, Belgium}
\author{Thomas Engl}
\email{Thomas.Engl@physik.uni-regensburg.de}
\affiliation{\RegensburgUniversity}
\author{Juan Diego Urbina}
\affiliation{\RegensburgUniversity}
\author{Klaus Richter}
\affiliation{\RegensburgUniversity}
\author{Peter Schlagheck}
\affiliation{\LiegeUniversity}
\maketitle
\section{Dimension of the Hilbert space}
The dimension of the Hilbert space for $N$ spin-$\frac{1}{2}$ particles on $L$ sites is determined by the number $\mathcal{N}$ of all possible occupations with no two particles with the same spin in the same site. This number is obtained by counting the number of possibilities to put $N$ indistinguishable spin-$\frac{1}{2}$ particles on $L$ sites with at most one particle per site, or equivalently to choose $N$ \emph{different} sites out of $2L$ (the factor $2$ accounts for the two possible choices for the spin) to put a particle on it without taking into account their order. This corresponds to an urn problem where from an urn with $2L$ balls, $N$ balls are drawn without putting them back and without considering the order, in which they are drawn. This standard problem in statistics has the well known solution of $2L \choose N$ possibilities. Thus there are
\begin{equation}
\mathcal{N}={2L \choose N}
\end{equation}
possible occupations, which is also the dimension of the Hilbert space.
\section{Derivation of Eq.~(13)}
In order to compute the number $\mathcal{N}_{\rm e}^{(0)}$ of possible occupations with the sites $1,\ldots,M$ empty or doubly occupied, consider first the possibilities to have $k$ of these $M$ sites doubly occupied and the remaining $M-k$ sites empty, {\it i.e.}~ to choose $k$ out of $M$ sites to be doubly occupied. This gives, in the same way as before, $M\choose k$ possibilities. However, for each of those, there are $2(L-M) \choose N-2k$ possibilities to distribute the remaining $N-2k$ spin-$\frac{1}{2}$ particles to the remaining $L-M$ sites, such that the total number of possible occupations with $k$ out of the $M$ flipped sites being doubly occupied and the remaining $M-k$ being empty is given by ${M \choose k}{2(L-M) \choose N-2k}$. Finally, one has to sum over all possible values of $k$ yielding
\begin{equation}
\mathcal{N}_{\rm e}^{(0)}=\sum\limits_{k=0}^{\min\left(M,\left\lfloor\frac{N}{2}\right\rfloor\right)}{M \choose k}{2(L-M) \choose N-2k}.
\label{eq:number_double-occupancies}
\end{equation}

\section{Derivation of Eqs.~(15) and (16)}
The number of possibilities to distribute $k$ particles onto the $L-M$ sites, for which the spins are not flipped intermediately, is given by $2(L-M) \choose k$. For each of these possibilities there are $2M \choose N-k$ possibilities to put the remaining $N-k$ particles onto the $M$ sites, which are subject to the spin flip. Thus, for fixed $k$, the number of possible occupations with $k$ particles on the $L-M$ sites, which are not affected by the intermediate spin flip is given by
\begin{equation}
\mathcal{N}_k={2(L-M) \choose k}{2M \choose N-k}.
\label{eq:number_non-flipped_fixed}
\end{equation}
From this, the number $\mathcal{N}_{\rm e}$ of possible occupations with an even number of particles in the $L-M$ sites, for which the spins are not flipped intermediately, is obtained by simply summing $\mathcal{N}_k$ over all possible even values of $k$,
\begin{equation}
\mathcal{N}_{\rm e}=\sum\limits_{k \text{ even}}\mathcal{N}_k=\sum\limits_{k=0}^{\left\lfloor\frac{N}{2}\right\rfloor}{2(L-M) \choose 2k}{2M \choose N-2k}.
\end{equation}
In the same way, the number $\mathcal{N}_{\rm o}$ of possible occupations with an odd number of particles in the $L-M$ sites, for which the spins are not flipped intermediately, is obtained by summing $\mathcal{N}_k$ over all possible odd values of $k$,
\begin{equation}
\mathcal{N}_{\rm o}=\sum\limits_{k \text{ odd}}\mathcal{N}_k=\sum\limits_{k=1}^{\left\lceil\frac{N}{2}\right\rceil}{2(L-M) \choose 2k-1}{2M \choose N-2k+1}.
\end{equation}
These quantities can now be used in order to evaluate
\begin{equation}
\begin{split}
\mathcal{P}_{M,T}^{\rm echo}({\bf n}&;\tau=0)= \\
1+\sum\limits_{\bf m}&\frac{p_{\rm cl}\left({\bf T}{\bf n},{\bf F}^{(M)}{\bf m};t\right)p_{\rm cl}\left({\bf m},{\bf n};t\right)}{P_{\rm cl}\left({\bf T}{\bf n},{\bf n};\tau\right)} \\
&\exp\left\{{\rm i}\pi\sum\limits_{l=1}^{L}\left[n_{l\downarrow}-n_{l\uparrow}+m_{l\uparrow}-\left({\bf B}^{(M)}{\bf m}\right)_{l\uparrow}\right]\right\}
\end{split}
\end{equation}
Assuming that $p_{\rm cl}({\bf n},{\bf m})$ is independent of ${\bf n}$ and ${\bf m}$, {\it i.e.}~$p_{\rm cl}=\mathcal{N}$, and using the conservation of particles implying
\begin{equation}
\begin{split}
\sum\limits_{l=1}^{L}&\left[n_{l\downarrow}-n_{l\uparrow}+m_{l\uparrow}-\left({\bf F}^{(M)}{\bf m}\right)_{l\uparrow}\right] \\
=&N-\sum\limits_{l=1}^{L}2n_{l\uparrow}+\sum\limits_{l=1}^{M}(m_{l\uparrow}-m_{l\downarrow})-\sum\limits_{l=M+1}^{L}(m_{l\downarrow}-m_{l\downarrow}) \\
=&N-\sum\limits_{l=1}^{L}2n_{l\uparrow}-\sum\limits_{l=1}^{L}m_{l\downarrow}+\sum\limits_{l=1}^{M}m_{l\uparrow}+\sum\limits_{l=M+1}^{L}m_{l\downarrow} \\
=&2\left(\sum\limits_{l=1}^{M}m_{l\uparrow}+\sum\limits_{l=1}^{L}n_{l\uparrow}\right)+\sum\limits_{l=M+1}^{L}\left(m_{l\uparrow}+m_{l\downarrow}\right)
\end{split}
\end{equation}
yields
\begin{equation}
\begin{split}
\mathcal{P}_{M,T}^{\rm echo}({\bf n};\tau=0)=&1+\frac{1}{\mathcal{N}}\sum\limits_{\bf m}(-1)^{\sum\limits_{l=M+1}^{L}\left(m_{l\uparrow}+m_{l\downarrow}\right)} \\
=&1+\frac{\mathcal{N}_{\rm e}-\mathcal{N}_{\rm o}}{\mathcal{N}}
\end{split}
\end{equation}
\section{Derivation of Eq.~(18)}
For $X={\rm id}$, the MBSE probability at $\tau=0$ is given by
\begin{equation}
\begin{split}
\mathcal{P}_{M,{\rm id}}^{\rm echo}({\bf n}&;\tau=0)=1+\sum\limits_{\bf m}\frac{f(N,M;\varphi)}{\mathcal{N}}
\end{split}
\end{equation}
with
\begin{equation}
f(N,M;\varphi)=\sum\limits_{\bf m}\exp\left[2{\rm i}\varphi\sum\limits_{l=1}^{M}\left(m_{l\downarrow}-m_{l\uparrow}\right)\right],
\end{equation}
which can be expressed in terms of the numbers $k_{\uparrow}=\sum_{l=1}^{M}m_{l\uparrow}$ of spin-up and $k_{\downarrow}=\sum_{l=1}^{M}m_{l\downarrow}$ of spin-down particles in the flipped sites as
\begin{equation}
f(N,M;\varphi)=\sum\limits_{k_{\uparrow}=0}^{N}\sum\limits_{k_{\downarrow}=0}^{N-k_{\uparrow}}\mathcal{N}_{k_{\uparrow},k_{\downarrow}}\exp\left[2{\rm i}\varphi\left(k_{\downarrow}-k_{\uparrow}\right)\right].
\end{equation}
Here,
\begin{equation}
\mathcal{N}_{k_{\uparrow},k_{\downarrow}}=\sum\limits_{\bf m}\delta_{\sum\limits_{l=1}^{M}m_{l\uparrow},k_{\uparrow}}\delta_{\sum\limits_{l=1}^{M}m_{l\downarrow},k_{\downarrow}}
\end{equation}
is the number of possible occupations with $k_{\uparrow}$ spin-up and $k_{\downarrow}$ spni-down particles in the flipped sites and can again be obtained from standard combinatorics.

First of all there are $M \choose k_{\uparrow}$ and $M \choose k_{\downarrow}$ possiblities to distribute the $k_{\uparrow}$ spin-up and $k_{\downarrow}$ spin-down particles, respectively, to the $M$ sites, for which the spins are flipped intermediately. Then, $N-k_{\uparrow}-k_{\downarrow}$ particles are left to be distributed on $L-M$ sites with two possibilities for the spin, which gives $2(L-M) \choose N-k_{\uparrow}-k_{\downarrow}$ possibilities. Thus, in total
\begin{equation}
\mathcal{N}_{k_{\uparrow},k_{\downarrow}}={M \choose k_{\uparrow}}{M \choose k_{\downarrow}}{2(L-M) \choose N-k_{\uparrow}-k_{\downarrow}}.
\end{equation}

\section{Derivation of Eq.~(21)}

Contrary to the peak heights, that are only determined by the symmetries of the system and therefore independent of the details of the mean-field Hamiltonian $H_{\rm MF}$, the calculation of the peak widths is sensitive to some more specific functional form of $H_{\rm MF}$. Nevertheless, the qualitative and some quantitative features of the peak widths can be obtained even without the precise knowledge of the classical Hamiltonian, as long as the interaction term is of the density-density form, as in Eq.~(3).

The starting point is again Eq.~(7). First, we note that the diagonal pairing gives a contribution that is fully independent of the mismatch $\tau$. Second, that in all pairings that give non-vanishing contributions to the transition probability, shown in Fig.~3, all trajectories involved have either ${\bf n}$ or ${\bf X}{\bf n}$ as final and/or initial configuration. For simplicity we will focus on ${\bf X}={\rm id}$, the other cases are easily calculated following the same idea.

For a given pair $(\gamma,\gamma^{\prime})$ of correlated trajectories, the action difference $\Delta=R_{\gamma}-R_{\gamma^{\prime}}$ comes from two sources. A mismatch $\Delta_{{\bf n}}$ of the initial/final configurations gives rise to the action differences shown in the main text. A second contribution $\Delta_{t}$ to the action differences comes from a mismatch $\tau$ of the propagation times and it is given by 
\begin{equation}
\Delta_{t}=R_{\bar{\gamma}}(t)-R_{\bar{\gamma}}(t^{\prime})
\end{equation}
where $\bar{\gamma}$ is a reference trajectory with initial and final configurations given by the average of the initial and final configurations of $\gamma,\gamma^{\prime}$. As usual, for large time differences the action difference will lead under averaging to large cancellations of oscillating functions. Only the neighborhood $t \simeq t^{\prime}$ matters then, and the expansion
\begin{equation}
\Delta_{t}\simeq \frac{\partial R_{\bar{\gamma}}(t) }{\partial t}\tau +{\cal O}(\tau^{2}), {\rm with \ \ }\tau=t-t^{\prime} 
\end{equation}
is justified. Now we use a well known property of the classical action function
\begin{equation}
\frac{\partial R_{\bar{\gamma}}(t) }{\partial t}=-E_{\gamma}(t)
\end{equation}
giving the numerical value of the mechanical energy as a function of the initial/final configurations and the propagation time $t$. With this we get the contributions to Eq.~(7) from the pairings (ii) and (iii), respectively, to be given by
\begin{eqnarray}
\label{eq:JD2}
P^{(\rm ii,iii)}(\tau)&=&\sum_{{\bf m},{\bf m}^{\prime}}\sum_{\gamma,\gamma^{\prime}}f_{{\bf m},{\bf m}^{\prime}}|A_{\gamma}|^{2}|A_{\gamma^{\prime}}|^{2}{\rm e}^{{\rm i}\Delta^{\gamma}_{\gamma^{\prime}}} \nonumber \\ &\times&{\rm e}^{{\rm i}E_{\gamma}({\bf n},{\bf m},t)\tau/\hbar}{\rm e}^{-{\rm i}E_{\gamma^{\prime}}({\bf m^{\prime}},{\bf n},t)\tau/\hbar},
\end{eqnarray}
where the function $f^{\gamma,\gamma^{\prime}}_{{\bf m},{\bf m}^{\prime}}$ accounts for the different possible action differences $\Delta$, as explained in the main text.

We will use now the fact that the classical, mean-field dynamics is assumed to be chaotic. First, the sums over trajectories in Eq.~(\ref{eq:JD2}) can be rewritten as phase space averages by means of the exact identity
\begin{equation}
\sum_{\gamma({\bf n},{\bf m},t)}F({{\bf m}})|A_{\gamma}|^{2}\propto\int_{0}^{2\pi}d\boldsymbol{\theta}F({\bf m}({\bf n},\boldsymbol{\theta},t),
\end{equation}
such that the full expression for $P(\tau)$ is then given by a classical correlation function between phase space functions $f$ and ${\rm exp}(i(E-E^{\prime})\tau/\hbar)$. Since in chaotic systems correlation functions $\langle F(0)G(t)\rangle$ display a fast exponential decay to their unconnected value $\langle F\rangle \langle G\rangle$ and these can be evaluated at the final or initial phase space points, we obtain finally Eq.~(21) of the main text,
\begin{equation}
\frac{P^{\rm (ii,iii)}\left({\bf n}^{\prime},{\bf n};\tau\right)}{P^{{\rm (ii,iii)}}\left({\bf n}^{\prime},{\bf n};0\right)}=\left|\int_{0}^{2\pi}{\rm e}^{\frac{{\rm i}}{\hbar}E\left({\bf n}^{\prime},{\boldsymbol \theta}\right)\tau}\frac{d^{2L}{\boldsymbol \theta}}{(2\pi)^{2L}}\right|^{2}.
\label{eq:JD1}
\end{equation}

\section{Universal features of the echo widths}

In general, the explicit evaluation of the echo widths from Eq.~(\ref{eq:JD1}) requires the explicit form of the mean-field Hamiltonian. Further progress can be made by making explicit the complex conjugation that define the modulus squared to see that the whole dependence of the widths is given by double phase space averages of the function
\begin{equation}
 {\rm e}^{\frac{{\rm i}}{\hbar}\left(H({\bf n}^{\prime},{\boldsymbol \theta})-H({\bf n}^{\prime},{\boldsymbol \theta}^{\prime})\right)\tau}.
\end{equation}
Therefore, for Hamiltonians where the interaction is of the density-density form and the dynamics is chaotic, the difference $H({\bf n}^{\prime},{\boldsymbol \theta})-H({\bf n}^{\prime},{\boldsymbol \theta}^{\prime})$ is fully independent of the interaction strength and the echo widths are then insensitive to $U$. This surprising result is fully confirmed by the simulations presented in Fig.~4 of the main text.

The dependence of the widths with the other microscopic parameters appearing in the hopping part of the Hamiltonian, $J$ and $\alpha$, can be estimated as follows. Once the interaction part of the Hamiltonian is canceled, each angular variable $\theta_{l}$ in Eq.~(\ref{eq:JD1}) appears only in the terms connecting the site $l$ with neighborhood sites. The explicit form of the integrand depends on the initial state, but each angular integration will in average contribute to the total width with a factor
\begin{equation}
\frac{1}{2\pi}\int_{0}^{2\pi}{\rm e}^{\frac{{\rm i}}{\hbar}J(\alpha)\tau\cos{\theta}}d\theta=J_{0}(J(\alpha)\tau/\hbar)
\end{equation}
where $J_{0}(x)$ is the zeroth order Bessel function. The width is then estimated by looking for the value of $\tau$ where the first zero of the product of Bessel functions appears. This will happen for $\tau\sim 1/{\rm Max}(J,\alpha)$, and gives the prediction of the width scaling roughly with the inverse of the maximum between the hopping $J$ and spin-orbit coupling $\alpha$. This prediction is also confirmed by the simulations shown in the main text.